\documentclass{PoS}
\def\al{\alpha}
\def\be{\beta}
\def\ga{\gamma}

\def\th{\theta}

\def\ps{\psi}

\def\cL{{\cal L}}

\def\frac#1#2{\textstyle{{{#1} \over {#2}}}}

\def\half{{\textstyle{1\over 2}}}
\def\lsim{\mathrel{\rlap{\lower4pt\hbox{\hskip1pt$\sim$}}
    \raise1pt\hbox{$<$}}}
\def\gsim{\mathrel{\rlap{\lower4pt\hbox{\hskip1pt$\sim$}}
    \raise1pt\hbox{$>$}}}

\def\lrDmu{\stackrel{\leftrightarrow}{D_\mu}}

\def\lrDbe{\stackrel{\leftrightarrow}{D_\be}}

\title{Lorentz Invariance Violation from String Theory}

\ShortTitle{LIV from String Theory}

\author{\speaker{Nikolaos  E. Mavromatos}%
         \thanks{This work has been supported in part by the European Science Foundation network
programme {\it Quantum Geometry and Quantum Gravity} and by the European Union through the FP6 Marie Curie Research and Training Network \emph{UniverseNet} (MRTN-CT-2006-035863).}\\
        King's College London, Department of Physics, Strand, London WC2R 2LS, U.K.\\
        E-mail: \email{nikolaos.mavromatos@kcl.ac.uk}}

\abstract{In this brief, and by no means complete, review I discuss situations in string theory, in which Lorentz Invariance Violation may occur in a way consistent with world-sheet conformal invariance, thereby leading to acceptable, in principle, string backgrounds. In particular, I first discuss spontaneous Lorentz violation in (non supersymmetric) open string field theory. Then, I move onto a discussion of gravity-induced modified dispersion relations in non-critical (Liouville) strings, in the sense of an induced Finsler-like geometry depending on both
coordinates and momenta, for string propagation in non-trivial space times (such as D-particle ``foamy situations''). I pay attention to explaining the appearance of bi-metric models from such string theories, which could serve as examples of alternative scenaria to dark matter. Finally, I make some comparisons with similar  developments in other contexts, such as critical strings in non-commutative space times, as well as deformed special relativities and theories with reduced Lorentz symmetry, advocated recently, where again Finsler geometry seems to come into play. In this latter respect, I put the emphasis on phenomenology and attempt to answer the question as to whether there is the possibility of experimental disentanglement of the various approaches. }

\FullConference{From Quantum to Emergent Gravity: Theory and Phenomenology\\
         June 11-15 2007\\
         Trieste, Italy}

\begin{document}

\section{Introduction}

Although classical general relativity entails  Lorentz invariance locally, and the latter symmetry survived many stringent experimental tests, especially the ones making use of high energy particle probes~\cite{coleman},
this symmetry may not be a true feature of Quantum Gravity.
In this talk I will discuss some instances in string theory where Lonretz symmetry could be broken. As I will argue, such situations are consistent with world-sheet conformal invariance, and hence, at least from a first-quantization point of view, are acceptable string theory backgrounds. The discussion will include strings in external electric fields, which are known to induce space-time non commutativity. I will also discuss some instances where second quantized open string field theory entails Lorentz Invariance Violations (LIV)~\cite{kosstring}. Moreover, I will describe how such LIV may arise in non-equilibrium situations in string theory, described in a first quantized framework by a version of the so-called non-critical (Liouville) string theory, in which the target time is identified with the (irreversible) Liouville mode~\cite{emn,emnrecoil}.

Among the topics to be discussed, are induced modified dispersion relations,
which may also characterize other approaches to quantum gravity, such as
deformed special relativities (DSR)~\cite{dsr}, which have been covered extensively in this meeting by other speakers.
I will attempt to make a comparison with such an approach, from the point of view of the induced coordinate and momentum dependent Finsler geometries~\cite{finslermetric,finsler}.

I will not be detailed in the phenomenology of the various approaches, as the subject is vast and has been covered at various sessions of this conference.
There are excellent reviews on the subject, see for instance ref.~\cite{jacobson}, where I refer the reader  for details. I will however mention very briefly some phenomenological issues, in particular those associated with consequences on Lorentz symmetry on CPT violation~\cite{poland}, which could help disentangle experimentally the various LIV models, in the sense that, as we shall see, LIV does not necessarily imply CPT Violation~\cite{sme}, and that there are certain forms of CPT Violation which could occur only in  specific ways of LIV~\cite{poland}.

It should be stressed that the review is from a personal perspective and is by no means complete. I do hope, however, that it captures some essential features of this rapidly expanding research field and communicates relevant and accurate information among the relevant communities.

The structure of the talk is as follows: in section 2 I discuss LIV from an open bosonic string field theory point of view. In section 3 I describe LIV in the context of a generic first quantized non-critical (Liouville) string, while in section 4 I discuss an example of the latter, that of a recoiling D-particle and the associated space time foam model. I pay particular emphasis on explaining how modified dispersion relations of particle probes emerge in such models, as a result of induced geometry deformations, described by space-time metrics that depend on both the coordinates and the momenta of the probe (Finsler type). In section 5 I compare the results with cases of strings in non commutative space times. In section 6 I make a brief comparison of the predictions of the D-particle model,
as far as modified dispersion relations and Finsler geometries are concerned,
with deformed special relativities, as well as a (deformed version~\cite{pope} of the) very special relativity with reduced symmetry advocated by Cohen and Glashow~\cite{cohen}. Finally section 7 states our conclusions and outlook.

\section{Spontaneous Lorentz Violation in Bosonic Open String Field Theory}

A fundamental way for breaking Lorentz symmetry in theories of quantum gravity is the {\it spontaneous breaking of Lorentz symmetry
(SBL)}~\cite{sme}, according to which
the ground state of a physical quantum system is characterized
by non trivial vacuum expectation values of certain tensorial
quantities, $\langle {\cal A}_\mu \rangle \ne 0$,
or $\langle {\cal B}_{\mu_1\mu_2\dots}\rangle \ne
0~$.
A concrete example of SBL may be provided by string field theory models of open bosonic strings~\cite{kosstring}. A generic state of an open string field can be expanded in terms of (an infinite series of) string modes as:
\begin{eqnarray}
&& |\Psi \rangle = \{ T(x_0) + A_\mu (x_0)a_{-1}^\mu + \frac{1}{\sqrt{2}}iB_\mu(x_0)a_{-2}^\mu + \frac{1}{\sqrt{2}}B_{\mu\nu}(x_0)a_{-1}^\mu a_{-1}^\nu + \nonumber \\ && \beta_1(x_0)b_{-1}c_{-1} + \dots + \frac{1}{2}i D_\mu (x_0)a_{-4}^\mu + \dots
\delta_3(x_0)b_{-1}c_{-3} + \dots \}| 0\rangle
\label{stringfield}
\end{eqnarray}
where $x_0$ denotes a generic space time point and $a_n^\mu$ are the appropriate creation operators, which upon acting on the (first quantized) open string vacuum state $| 0 \rangle$, create the various excitation modes: scalars (tachyons) with amplitude $T(x_0)$, which are characteristic of the open bosonic (in general broken supersymmetric) string,
vectors, with amplitudes $A_\mu, \dots$, tensors with amplitudes $B_{\mu\nu}, \dots$ {\it etc.}. Ghost fields, with creation operators $b_n, c_n$ are also included, which arise from fixing the gauge invariances of the string state.

The open string field theory action is cubic in the field $\Psi$.
Upon expanding about a squeezed state background $\Psi_B$, $\Psi = \Psi_B + \Delta $, as appropriate for our discussion on Lorentz violating vacua in strings~\cite{kosstring}, one obtains for the string field theory action $I(\Psi)$:
\begin{equation}
I(\Psi) = \frac{1}{2\alpha '}\int \Psi_B \star Q\Psi_B + \int \frac{g}{3}\Psi_B \star \Psi_B \star \Psi_B + \frac{1}{2\alpha '} \int \Delta \star Q_B \Delta + \int \frac{g}{3} \Delta \star \Delta \star \Delta~.
\label{actionft}
\end{equation}
In the above equation, $\star$ denotes the appropriate gauge invariant inner product of open string field theory, $\alpha '$ is the Regge slope, and the suffix $B$ denotes background, whilst $g$ is the ghost number. The quantity $Q$ is the nilpotent ($Q^2=0$) BRST operator, and $Q_B$ is defined by its action on a generic string state $\Phi$ in the background $\Psi_B$:
$ Q_B \Phi = Q\Phi + g\alpha ' [\Psi_B \star \Phi - (-1)^{g(\Phi)}\Phi \star \Psi_B] $.

As becomes evident from (\ref{actionft}), in such models, there are cubic terms in an effective low-energy (target-space) Lagrangian involving the
tachyonic scalar field $T$, that characterizes the bosonic string vacuum, and invariant products of higher-tensor fields $ {\cal B}_{\mu_1 \dots \mu_n} $
that appear in the mode expansion of a string field:
\begin{equation}
T {\cal B}_{\mu_1 \dots \mu_n}{\cal B}^{\mu_1 \dots \mu_n}~.
\label{TBB}
\end{equation}

The negative mass squared tachyon field, then, acts as a Higgs field in
such theories, acquiring a vacuum expectation value, which, in turn, implies
non-zero vacuum expectation values for the tensor fields ${\cal B}$, leading in this way to energetically preferable configurations that are Lorentz Invariance Violating (LIV). From the point of view of string theory landscape these are perfectly acceptable vacua~\cite{kosstring}, given that they respect world-sheet conformal invariance of the first quantized string theory.

An effective target-space field theory framework to discuss the phenomenology
of such LIV theories is the so-called Standard Model Extension (SME)~\cite{sme}. For our purposes in this section, it suffices to simply give an example of SME effects on phenomenology of particle physics, by considering the
SME \emph{Modified Dirac Equation} for spinor fields $\psi$, representing leptons and quarks with
charge $q$:\begin{equation}\label{diracsme}
\left( i\gamma^\mu D^\mu - M -  a_\mu \gamma^\mu
- b_\mu \gamma_5
\gamma^\mu  -
\frac{1}{2}H_{\mu\nu}\sigma ^{\mu\nu} +
ic_{\mu\nu}\gamma^\mu D^\nu + id_{\mu\nu}\gamma_5\gamma^\mu D^\nu
\right)\psi =0~,
\end{equation}where $D_\mu = \partial_\mu - A_\mu^a T^a - qA_\mu$ is an
appropriate gauge-covariant derivative. The non-conventional terms
proportional to the coefficients $a_\mu,~  b_\mu,~ c_{\mu\nu},~
d_{\mu\nu},~ H_{\mu\nu}, \dots $, stem from the corresponding local
operators of the effective Lagrangian which are phenomenological at
this stage. The set of terms pertaining to $a_\mu~, b_\mu$ entail
CPT and Lorentz Violation, while the terms proportional to
$c_{\mu\nu}~, d_{\mu\nu}~, H_{\mu\nu}$ exhibit Lorentz Violation
only. It should be stressed that, within the SME framework, as is also
the case with the decoherence approach to Quantum Gravity (QG)~\cite{poland}, CPT violation does \emph{not
necessarily} imply  mass differences between particles and
antiparticles.

Some remarks are now in order, regarding the form and
order-of-magnitude estimates of the Lorentz and/or CPT violating
effects. In the approach of \cite{sme} the SME
coefficients have been taken to be constants. Unfortunately there is
not yet a detailed microscopic model available, which would allow
for concrete predictions of their order of magnitude.
Theoretically, the (dimensionful, with dimensions of energy) SME
parameters can be bounded by applying renormalization group and
naturalness assumptions to the effective local SME Hamiltonian,
which leads to bounds on $b_\mu$ of order $10^{-17}~{\rm
GeV}$. At present all SME parameters should be considered as
phenomenological and to be constrained by experiment.
In general, however, the constancy of the SME coefficients may not
be true. In fact, in certain string-inspired or stochastic models of
space-time foam that violate Lorentz symmetry~\cite{poland,bms}, the coefficients
$a_\mu, b_\mu ...$ are probe-energy ($E$) dependent, as a result of
back-reaction effects of matter onto the fluctuating space-time.
Specifically, in stochastic models of space-time foam, one may find~\cite{bms}
that on average there is no Lorentz and/or CPT violation, i.e., the
respective statistical v.e.v.s (over stochastic space-time
fluctuations)
 $\langle a_\mu~, b_\mu \rangle = 0~,$ but this is not true for
 higher order correlators of these quantities (fluctuations), i.e.,
$\langle a_\mu a_\nu \rangle \ne 0,~
 \langle b_\mu a_\nu \rangle \ne 0,~  \langle b_\mu b_\nu \rangle \ne 0~, \dots
 $. In such a case the SME effects will be
much more suppressed, since by dimensional arguments such
fluctuations are expected to be at most of order $E^4/M_P^2$,
and hence much harder to detect.

\section{Non-Critical String theory as an alternative to Landscape and LIV}

\begin{figure}[ht]
\begin{center}
\includegraphics[width=7cm]{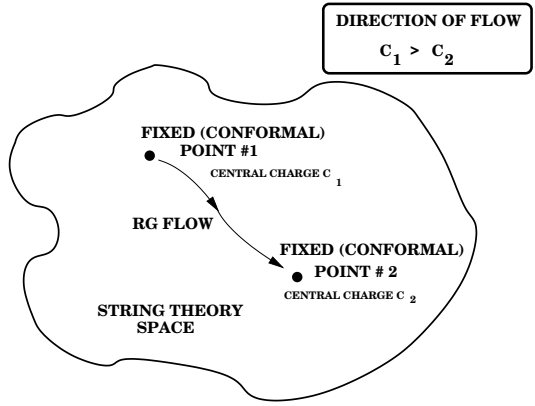}
\end{center}
\caption{\it
A schematic view of string theory space, which is an infinite-dimensional
manifold endowed with a (Zamolodchikov) metric. The dots denote conformal
string backgrounds. A non-conformal string flows (in a two-dimensional
renormalization-group sense)  from one fixed point to another, either of
which could be a hypersurface in theory space.  The direction of the flow
is irreversible, and is directed towards the fixed point with a lesser
value of the central charge, for unitary theories, or, for general
theories, towards minimization of the degrees of freedom of the system.}
\label{fig:flow}
\end{figure}

Critical strings, in a first quantized form~\cite{gsw}, endow world-sheet conformal invariance. As a result, from a target space time view point the strings propagate in classical space time backgrounds that obey equations of motion, derived from an effective action, and thus they describe by construction {\it equilibrium} situations in field theory.
On the other hand, non-critical (Liouville) strings~\cite{ddk} correspond to $\sigma$-models deformed by world-sheet vertex operators that are non conformal. As a result, the corresponding target-space backgrounds, which the string propagates on, are {\it off-shell}, that is they do not obey equations of motion, although the corresponding world-sheet beta functions are proportional
to off-shell variations of the string effective action. This latter result stems from a generic property of stringy $\sigma$-models, according to which the world-sheet renormalization group beta function, $\beta^i(g)$, expressing the running of the ``renormalized'' background $g^i(\mu) $ with the world-sheet renormalization scale $\mu$, is always proportional to the variations with respect to $g^i$ of an off-shell scalar function ${\cal S}$ in the space of backgrounds $\{ g^i \}$, which thus plays the r\^ole of a target space action~\cite{osborn}:
\begin{equation}\label{offshellbeta}
     \beta^i (g) = {\cal G}^{ij}\frac{\delta {\cal S}}{\delta g^j}
\end{equation}
where ${\cal G}^{ij}$ is the inverse of Zamolodchikov metric~\cite{zam} in the theory space of strings, $\{ g^i \}$, which is related~\cite{zam,osborn} to world-sheet short distance divergencies of the two point function of the vertex operators for the backgrounds $g^i$: ${\cal G}_{ij} = {\rm Lim}_{z \to 0} z^2{\overline z}^2\langle V_i(z,{\overline z}) V_j (0,0)\rangle $.

In this sense, non-critical strings represent {\it non equilibrium} situations in string theory, and in fact have been used~\cite{emninfl,emnw} to discuss an approach to equilibrium in cosmology, in an attempt to explain the smallness of the observed current-epoch cosmological constant (or, more accurately, dark energy), by viewing it as a relaxation phenomenon (see fig.~\ref{fig:flow}).
The framework may be viewed as providing alternatives to landscape scenarios in string theory~\cite{emn,emnw}.
Crucial to the above interpretation was the identification~\cite{emn} of time with the Liouville mode of non-critical strings~\cite{ddk}, which is allowed in certain {\it supercritical} string models~\cite{aben}, described by $\sigma$-models whose central charge exceeds the critical value. We note that recently, supercritical strings, especially from the cosmological point of view, attracted some attention, either from a theoretical point of view, in an attempt to discuss the initial value problem of our Universe and the associated cosmological instabilities~\cite{recentssc}, or from a purely phenomenological view point, as modifying astroparticle physics constraints on interesting particle physics models, such as supersymmetric~\cite{lmn}.

In this presentation we shall consider non-critical strings~\cite{emn} as providing~\cite{aemn} situations in which Lorentz symmetry of the string vacuum may be broken, in the sense of leading to a non-trivial {\it vacuum refractive index}, i.e. modified dispersion relations, for photons. In fact, to our knowledge, this was the first instance where modified dispersions have been proposed in the context of concrete Lorentz violating approaches to quantum gravity. Subsequently, many other proposals have been made, which entailed non-standard dispersion relations for a variety of reasons that I will not discuss here. For the purposes of this talk, I will single out the so-called Deformed Special Relativity (DSR) class of models, in which such modified dispersion relations are a consequence of the deformed Lorentz symmetries that
leave invariant the characteristic energy scale (``Planck'') of the theory~\cite{dsr}. We shall compare the Liouville string results with DSR later on in our presentation.

At present, we consider it as instructive to review first the basics of the non-critical-string formalism that lead to such results. To this end,
let one consider a $\sigma$-model action deformed by
a family of
vertex operators $V_i$, corresponding to
`couplings' $g^i$, which represent \emph{non-conformal} background space-time
fields from the massless string multiplet, such as
gravitons, $G_{\mu\nu}$, antisymmetric tensors, $B_{\mu\nu}$,
dilatons $\Phi$, their supersymmetric partners, \emph{etc.}:
\begin{equation}
S=S_{0}\left( X\right) +\sum_{i}g^{i}\int d^{2}z\,V_{i}\left(
X \right)~,
\label{action}
\end{equation}
where $S_0$ represents a conformal $\sigma$ model describing
an equilibrium situation.
The non-conformality of the background means
that the pertinent $\beta^i$ function $\beta^i \equiv dg^i/d{\rm ln}\mu
\ne 0$, where $\mu$ is a world-sheet renormalization scale.
Conformal invariance would imply restrictions on the
background and couplings $g^i$, corresponding to
the constraints $\beta^i = 0$, which are equivalent to equations
of motion derived from a target-space effective action for the corresponding
fields $g^i$.  The entire low-energy
phenomenology and model building of critical string theory is based on such
restrictions~\cite{gsw}.

In the non-conformal case $\beta^i \ne 0$, the theory is in need of
dressing by the Liouville field $\phi$ in order to restore conformal
symmetry~\cite{ddk}. The field $\phi $ acquires dynamics through the
integration over world-sheet covariant metrics in the path integral, and
may be viewed as a local dynamical scale on the world sheet~\cite{emn}.
If the central charge of the (supersymmetric)  matter theory is $c_{m}>25
(9)$ (i.e., supercritical~\cite{aben}), the signature of the kinetic term of the
Liouville coordinate in the dressed $\sigma$-model is opposite to that of
the $\sigma$-model fields corresponding to the other target-space
coordinates.  As mentioned previously, this opens the way to the important
step of interpreting the Liouville field physically by identifying its
world-sheet zero mode $\phi _{0}$ with the target time in supercritical
theories~\cite{emn}.  Such an identification emerges naturally from the
dynamics of the target-space low-energy effective theory by minimizing the
effective potential~\cite{gravanis}.

The action of the Liouville mode $\phi$ reads~\cite{ddk}:
\begin{equation}
S_L = S_{0}\left( X\right)
+ \frac{1}{8\pi} \int_\Sigma d^2\xi \sqrt{\widehat \gamma}
[ \pm (\partial \phi)^2 - QR^{(2)}\phi] + \int_\Sigma d^2\xi
\sqrt{\widehat \gamma} g^i(\phi)V_i(X)~,
\label{liouvilleact}
\end{equation}
where ${\widehat \gamma}$ is a fiducial world-sheet metric, and
the plus (minus) sign in front of the kinetic term of the
Liouville mode pertains to subcritical (supercritical) strings.
The dressed couplings $g^i(\phi)$  are obtained by the following procedure:
\begin{equation}\label{dressing}
\int d^{2}z\,\,g_{i} \,V_{i}\left(X\right) \rightarrow \int
d^{2}z\,\,g_{i}(\phi)\,\,e^{\alpha _{i}\phi }\,V_{i}\left(X\right)~,
\end{equation}
where $\alpha_i$ is the
``gravitational'' anomalous dimension. If the original non-conformal
vertex operator
has anomalous scaling dimension $\Delta_i -2$ (for closed strings,
to which we restrict
ourselves for definiteness), where $\Delta_i$ is the conformal dimension,
and the central charge surplus
of the theory is $Q^2 = \frac{c_m - c^*}{3} > 0 $ (for bosonic strings
$c^*=25$, for superstrings $c^*=9$), then the condition
that the dressed operator is marginal on the world sheet implies the
relation:
\begin{equation}\label{lad}
\alpha_i (\alpha_i + Q) = 2 - \Delta_i~.
\end{equation}
Imposing appropriate boundary conditions in the limit
$\phi \to \infty$~\cite{ddk}, the acceptable solution is:
\begin{equation}\label{solutions}
\alpha_i = -\frac{Q}{2} + \sqrt{\frac{Q^2}{4} + 2 - \Delta_i}~.
\end{equation}
The gravitational dressing is trivial for marginal couplings,
$\Delta_i=2$, as it should be.
This dressing applies also to higher orders in the
perturbative $g^i$ expansion. For instance,
at the next order, where the deviation from marginality
in the deformations of the undressed $\sigma$model
is due to the operator product expansion coefficients
$c^i_{jk}$ in the $\beta^i$ function, the Liouville-dressing
procedure implies the replacement~\cite{schmid}:
\begin{equation}
g^i \qquad \to \qquad g^ie^{\alpha_i\phi} + \frac{\pi~\phi}{Q \pm
2\alpha_i}c^i_{jk}g^jg^ke^{\alpha_i \phi} ,
\label{dressing2}
\end{equation}
in order for the dressed operator to become marginal to this order
(the $\pm$ sign originates in (\ref{liouvilleact})).

In terms of the Liouville renormalization-group scale, one
has the following equation relating Liouville-dressed
couplings $g^i$ and $\beta$ functions in the non-critical string case:
\begin{equation}
{\ddot  g}^i + Q{\dot g}i = \mp\beta^i(g_j)~,
\label{liouvilleeq}
\end{equation}
where the $-~(+)$ sign in front of the $\beta$-functions
on the right-hand-side applies to super (sub)critical strings,
the overdot denotes differentiation with respect to the
Liouville zero mode and $\beta^i$ is the world-sheet renormalization-group
$\beta$ function satisfying the gradient flow relation (\ref{offshellbeta}) (but with the renormalized couplings replaced
by the Liouville-dressed ones as defined by the procedure in
(\ref{dressing}), (\ref{dressing2})).
Formally, the $\beta^i$ of the r.h.s.\ of (\ref{liouvilleeq})
may be viewed as
power series in the (weak) couplings $g^i$.
The covariant (in theory space)
${\cal G}_{ij}\beta^j$ function may be expanded as:
\begin{equation}
{\cal G}_{ij}\beta^j =
\sum_{i_n} \langle V_i^L V_{i_1}^L \dots V_{i_n}^L \rangle_\phi g^{i_1}
\dots g^{i_n}~,
\label{betaexp}
\end{equation}
where $V_i^L$ indicates Liouville dressing  \`a la (\ref{dressing}),
$ \langle \dots \rangle_\phi = \int d\phi d\vec{r}~{\rm exp}(-S(\phi, \vec{r}, g^j))$
denotes a functional average including Liouville integration, and
$S(\phi, \vec{r}, g^i)$ is the Liouville-dressed $\sigma$-model
action, including the Liouville action~\cite{ddk}.

In the case of stringy $\sigma$ models,
the diffeomorphism invariance of the target space results in the
replacement of (\ref{liouvilleeq}) by:
\begin{equation}
  {\ddot g}^i + Q(t){\dot g}^i = \mp{\tilde \beta}^i ,
\label{liouvilleeq2}
\end{equation}
where the ${\tilde \beta}^i$ are the Weyl anomaly coefficients of the
stringy $\sigma$ model in the background $\{ g^i \}$, which differ
from the ordinary world-sheet renormalization-group $\beta^i$ functions
by terms of the form:
\begin{equation}
{\tilde \beta}^i = \beta^i + \delta g^i
\end{equation}
where $\delta g^i$ denote transformations of the
background field $g^i$ under infinitesimal general coordinate
transformations, e.g., for gravitons~\cite{gsw}
${\tilde \beta}^G_{\mu\nu} =
\beta^G_{\mu\nu} + \nabla_{(\mu} W_{\nu)}$,
with $W_\mu = \nabla _\mu \Phi$,
and $\beta^G_{\mu\nu} = R_{\mu\nu}$ to order $\alpha '$
(one $\sigma$-model loop).

The set of equations (\ref{liouvilleeq}),(\ref{liouvilleeq2})  defines the
\emph{generalized conformal invariance conditions}, expressing the
restoration of conformal invariance by the Liouville mode.  The solution
of these equations, upon the identification of the Liouville zero mode
with the original target time, leads to constraints in the space-time
backgrounds~\cite{emn,gravanis}, in much the same way as the conformal
invariance conditions $\beta^i = 0$ define consistent space-time
backgrounds for critical strings~\cite{gsw}.  It is important to
remark~\cite{emninfl, emn, szabo} that the equations (\ref{liouvilleeq2}) can be
derived from an action.  This follows from general properties of the
Liouville renormalization group, which guarantee that the appropriate
Helmholtz conditions in the string-theory space $\{ g^i \}$ for the
Liouville-flow dynamics to be derivable from an action principle are
satisfied. It is equations of the form (\ref{liouvilleeq}),(\ref{liouvilleeq2}) that characterize the non-equilibrium Liouville string cosmologies~\cite{emnw}, mentioned previously, which replace the standard Einstein-Friedmann equations.

Now we come to our main point, namely the emergence of modified dispersion relations for particle probes in the non-critical string framework.
The structure of eq.~(\ref{dressing2}) suggests that the
effects of the non-criticality are quite complicated
in general. However, the form of the Liouville time dependence
implies that one of the physical effects of the
non-criticality is a modification
of the time-dependence of the Liouville dressed $g ^i(\phi = -t)$, which may be
described to order $O(g^2)$ by the following approximation
to (\ref{dressing2}):
\begin{equation}
   g ^i (t) \sim g^i e^{ i(\alpha _i + \Delta \alpha _i)t}
\label{estimate}
\end{equation}
where the $\Delta \alpha _i $ depend on the $c^i_{jk}$,
which encode the interactions with quantum-gravity
fluctuations in the space-time background.
We are now ready to discuss the issue of wave dispersion.
From a target-space point of view, the $g^i$ may be viewed
as the Fourier transforms, {\it i.e.}, the
polarization tensors, of target-space background fields,
and the vertex operators are wave
operators~\footnote{Here the concepts of Fourier
transforms and plane waves should be understood
as being appropriately generalized to curved target spaces,
with the appropriate geodesic distances taken into account.
For our purposes below, the details of this will not be relevant.
We shall work with macroscopically-flat space-times, where
the quantum-gravity structure appears
through quantum fluctuations of the vacuum, leading
simply to non-criticality of the string, in the sense
of non-vanishing $\beta ^i$ functions.}.
For instance, for a deformation corresponding to a
scalar mode $T(X)$, one can write
\begin{equation}
   \int d^2 z \sqrt{\gamma} g^i V_i \equiv
\int d^2 z \sqrt{\gamma} \int d^D k e^{i k_M X^M(z,{\bar z})}
{\tilde T}(k)
\label{scalar}
\end{equation}
where the summation over $i$ includes target-space integration
over $k$, on a $D$-dimensional Euclidean space.
For massless string modes, as opposed to higher modes,
$\Delta_i-2 = k^Mk_M \equiv |\vec k |^2$.
For our purposes, it suffices to consider the case of an
almost flat space time with small quantum-gravity corrections
coming from the interactions of massless low-energy modes
with the environment of Planckian string states,
implying that the string is close to a fixed point
for which $Q^2 = c[g^*]-25 = 0$. This is the case in
the backgrounds of interest to us, where the deviations from the conformal point are small.
In such a case, we see from (\ref{solutions}) that
$\alpha _i \sim |\vec{k}|$.

Our basic working hypothesis, which has been confirmed
explicitly in the two-dimensional string black hole
example~\cite{emn},
is that the matter deformations in such a black-hole background
are not exactly marginal unless the couplings to
non-propagating Planckian global modes
are included in the analysis. This implies that the set of the
operator product expansion coefficients
$c^i_{jk}$ in (\ref{dressing2}) includes non-zero couplings between
massless and Planckian modes in the presence of non-trivial
metric fluctuations, say of black-hole type, and more generally
in the presence of structures with a space-time
boundary. These couplings arise
in string theory from the infinite set of
stringy gauge $W_\infty$ symmetries. They imply that the massless modes constitute an {\it open
system}, and that they lose information to these non-propagating
higher-level modes which are not detected
in local scattering experiments, inducing apparent
decoherence~\cite{emn}.

The order of magnitude of such couplings is
not known at present, and precise calculations would require
a full second-quantized string theory.
As we shall discuss in the next section,
one may have situations in non-critical string models, in
which there is {\it minimal} suppression of the quantum gravity effects by a single power of the quantum gravity scale (say, Planck mass although in string theory this may be different from the string scale $M_s$). In such a case
one may assume that
\begin{equation}
\Delta \alpha _i \sim \eta {|\vec{k}|^2 \over M_P}
\label{etadef}
\end{equation}
with $\eta$ a dimensionless quantity, which
parametrizes our present
ignorance of the quantum structure of space time.
For the purposes of the present work,
we restrict ourselves to non-critical strings on
fixed-genus world sheets, in which case $\eta$ in (\ref{etadef})
is real. This should be viewed as describing only part of the
quantum-gravity entanglement, namely that associated with the
presence of global string modes~\cite{emn}. The full
string problem
involves a summation over genera, which in turn implies
{\it complex} $\eta$'s, arising from the appearance
of imaginary parts in Liouville-string correlation functions
on re-summed world sheets, as a result of instabilities pertaining to
microscopic black-hole decay in the quantum-gravity
space-time foam~\cite{emn}. Such imaginary parts
in $\eta$ will produce frequency-dependent
attenuation effects in the amplitudes of quantum-mechanical
waves for low-energy string modes. This type
of attenuation effect leads to
decoherence of the type appearing in the density-matrix
approach to measurement theory~\cite{emn}.
For our purpose of deriving bounds on the possible
accuracy of distance measurements, however,
such attenuation effects need not be taken into account, and
the simple entanglement formula (\ref{etadef}), with real $\eta$,
will be sufficient.

The value of $\eta$ depends in general on the type of massless
field considered. In particular, in the case of photons
$\eta$ is further constrained
by target-space gauge invariance, which restricts
the structure of the relevant $\sigma$-model $\beta$ function.
With this caveat in mind, (\ref{etadef}) represents a maximal
estimate of $\Delta \alpha _i$, compatible
with the generic structure of perturbations of the $\sigma$-model
$\beta$ functions for closed strings.

The equations (\ref{dressing2}), (\ref{scalar}) and (\ref{etadef})
indicate that the dressed $\sigma$-model deformation (\ref{dressing})
corresponds to waves of the form\footnote{Note that eq.~(\ref{wave})
is consistent with target-space gauge
invariance, since it may describe
one of the polarization states of the photon, and
the transversality condition on the photon
polarization tensor, $k^\mu A_\mu = 0$, can be
maintained.}
\begin{equation}
 e^{i |{\vec k}| t + i{\vec k}. {\vec X}  +
 i \eta {|\vec{k}|^2 \over M_P} t }
\label{wave}
\end{equation}
The corresponding modified dispersion relation
\begin{equation}
E \! \sim \! |\vec{k}| + \eta |\vec{k}|^2 /M_P,
\label{dispersion}
\end{equation}
which was dictated by the overall conformal invariance
of the dressed theory,
implies that massless particles propagate in the quantum-gravity
`medium' with a (group) velocity (in units of $c=1$) that is effectively
energy-dependent:
\begin{equation}
v \! \sim \! 1+ \eta E/M_P~.
\label{refindex}
\end{equation}

In the next section we shall discuss concrete models~\cite{emnmdr}, in which the induced non-criticality will arise as a result of microscopic interactions of the
massless string excitation, say photon, with a target-space point-like defect, a so-called D-particle, in the modern approach to string theory involving Dirichlet brane defects~\cite{polchinski}. In such models, the modified dispersion arises as a result of a non-Minkowskian, {\it momentum} dependent,
local deformation of the space-time metric, due to the ``recoil'' of the defect during its scattering with the massless string mode. One then obtains {\it effective } refractive indices {\it in vacuo} of the form (\ref{refindex}) with
$-1 < \eta < 0 $ and thus there is only {\it subluminal} propagation of photons in the D-particle medium.

\section{Gravity-Induced Modified Dispersion Relations in Non-Critical Strings and Finsler Geometry}

\begin{figure}[ht]
\centering
\includegraphics[width=6.5cm]{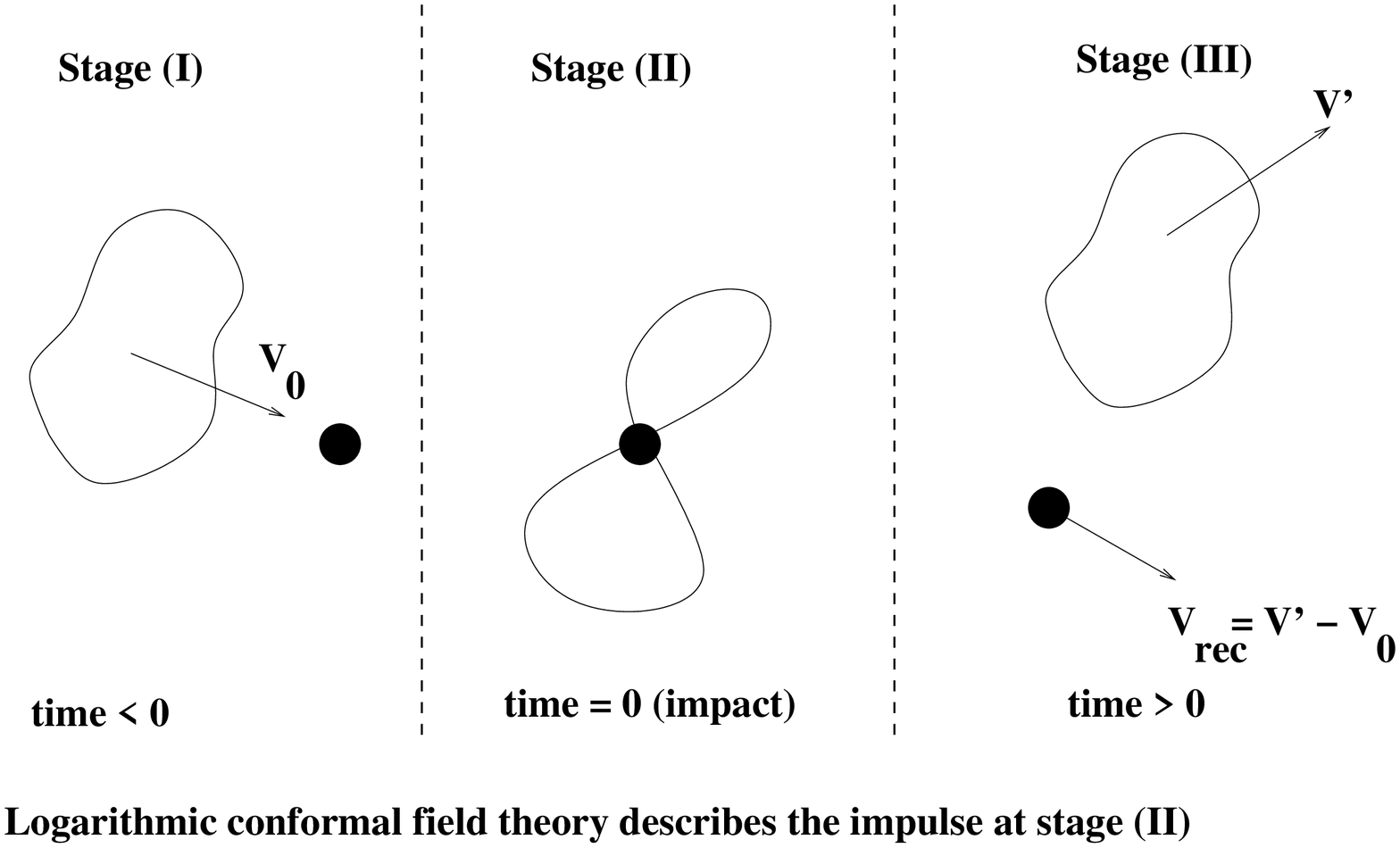} \hfill
\includegraphics[width=6.5cm]{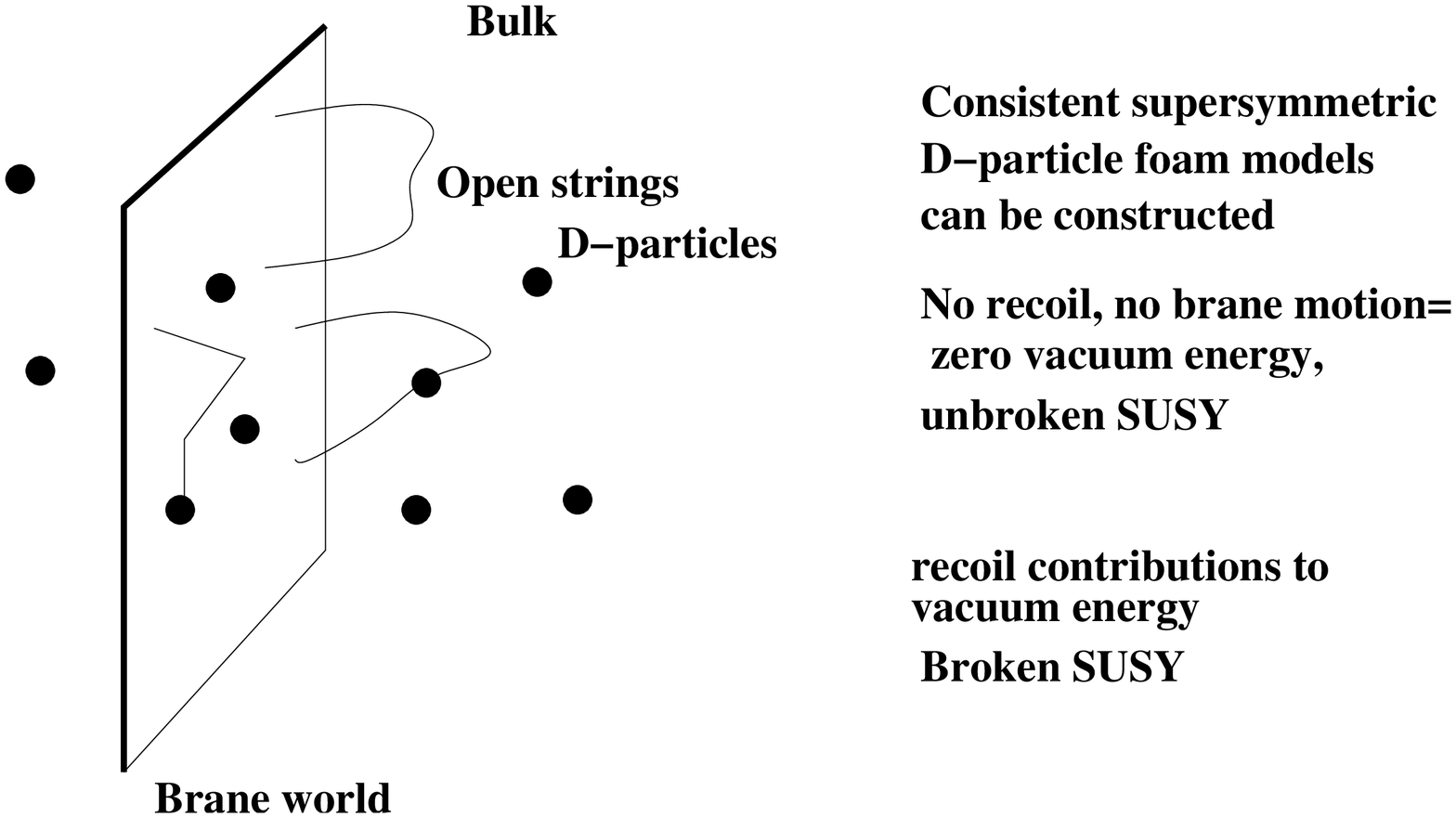}
\caption{Schematic representation of the capture/recoil process of a string state by a D-particle defect for closed (left) and open (right)
string states, in the presence of D-brane world. The
presence of a D-brane is essential due to gauge flux conservation, since an isolated D-particle cannot exist. The
intermediate composite state at $t=0$, which has a life time within the stringy uncertainty time interval $\delta t $, of the order of the string length, and is described by world-sheet logarithmic conformal field theory, is responsible for the distortion of the surrounding space time during the scattering, and subsequently leads to Finsler-type induced metrics (depending on both coordinates and momenta of the string state) and modified dispersion relations for the string propagation.}
 \label{fig:recoil}
 \end{figure}

We consider the situation depicted in fig.~\ref{fig:recoil}, which represents our world as a three brane, propagating in a bulk higher-dimensional space time, ``punctured'' by D-particle (point-like Dirichlet brane) defects~\footnote{The reader's attention is called here upon the necessary presence of a D-brane world when D-particles are embedded in the bulk space, given that isolated D-particles cannot exist~\cite{strominger} due to the requirement of conservation of the U(1) gauge fluxes that characterize D-branes~\cite{polchinski}. In general, in string/brane theory higher-rank $Dp$-branes (i.e. with p longitudinal directions) provide endpoints for lower-$p$ branes, and the latter govern the dynamics of the former.}. Such configurations are acceptable (super)string backgrounds, given that when the D-brane and D-particle configurations are static, one can construct a consistent supersymmetric ground state for the string with vanishing energy~\cite{emw}. Matter stringy excitations on the brane world, represented by open strings with their ends attached on the brane, as well as closed strings propagating on {\it both} the bulk and the brane, and representing excitations from the string gravitational multiplet (i.e. gravitons, dilatons, antisymmetric Kalb-Ramond tensors and their supersymmetric partners) can interact non trivially with the D-particles in the sense of the ``capture'' process indicated on the figure. According to this picture, there is an intermediate composite string-D-particle state~\cite{kmw}, which lives within a stringy time uncertainty interval $\delta t$, whose recoil (in order to maintain
energy-momentum conservation) distorts the surrounding space time~\cite{emnrecoil,emnmdr}, deforms it from the initially assumed Minkowski, and thus the string travels on a curved background. The deformation is {\it local} in the sense that it pertains only to the neighborhood of the D-particle defect and decays far away from it.

Logarithmic conformal field theory methods on the world sheet of the string have been used~\cite{kmw,szabo} in order to study the deformations of the $\sigma$-model produced as a result of the formation of the intermediate string-D-particle state and the subsequent recoil process, depicted in  fig.~\ref{fig:recoil}.
We shall omit the details here, since the reader may find them on the relevant works in the literature, and outline instead only the main results. The pertinent deformation on the open-string $\sigma$-model boundary $\partial \Sigma$, describing open string excitations of the D-particle as a result of its recoil,
is given by the impulse operator in a non-relativistic approximation, valid for the case of heavy D-particles we consider here~\cite{kmw,emnrecoil}:
\begin{equation}
V_{\rm rec} = \int_{\partial \Sigma} u_i X^0 \Theta_\varepsilon (X^0) \partial^n X^i ~, \qquad \Theta_\varepsilon (X^0) \equiv {\rm Lim}_{\varepsilon \to 0^+} \frac{1}{i}\int_{-\infty}^{+\infty} d\omega \frac{e^{i\omega X^0}}{\omega - i\varepsilon}
\label{recoilop}
\end{equation}
where $\partial^n$ denotes normal derivative on the world sheet. The fields $X^0$ obey Neumann (free) boundary conditions, whilst the fields $X^i$ ($i$ a spatial target-space index, $i=1, \dots D-1$ for $D$-dimensional embedding space times) satisfy Dirichlet (fixed) boundary conditions on the world sheet~\cite{polchinski}, as appropriate for a description of open string excitations attached to a D-brane (in this case a point-like one, the D-particle).

Logarithmic world-sheet algebra considerations~\cite{kmw,szabo} require the presence of a companion operator to close the algebra,
\begin{equation}
V_{\rm comp} = \int_{\partial \Sigma} \varepsilon y_i \Theta_\varepsilon (X^0) \partial^n X^i ~,
\label{recoilopII}
\end{equation}
where $y_i $ is a collective spatial coordinate of the D-particle. Physically
$\varepsilon y_i$ indicates the uncertainty in the spatial coordinate of the D-particle. This, then implies that the time uncertainty (in this approximation) is~\cite{kmw} $\Delta X^0 \equiv \Delta t \sim \frac{1}{\varepsilon} $, which describes the duration of the scattering, not to be confused with the life time of the intermediate composite string state of fig.~\ref{fig:recoil}, which is much smaller, of the order of the string length $\ell_s$~\cite{horizon}~\footnote{As discussed in \cite{kmw}, the presence of both pairs of logarithmic operators (\ref{recoilop}), (\ref{recoilopII}) are responsible for reproducing the correct stringy uncertainty principle between spatial coordinates and momenta, as well as all the other (stringy) uncertainties. Due to the presence of the D-particle, the
total energy of the composite state is $M_s/g_s$, to a first approximation (for probe energies much smaller than $M_s/g_s$), and hence its life time $\delta t \sim g_s /M_s = g_s \ell_s$~.}. It turns out that, for reasons pertaining to the closure of the logarithmic algebra~\cite{kmw} the limit $\varepsilon \to 0^+$ cannot be taken independently of the world sheet renormalization scale ${\rm ln}\Lambda$, where $\Lambda$ is the world-sheet area. In fact, one must have
\begin{equation}
\varepsilon ^{-2} \sim {\rm ln}\Lambda
\label{cutoffepsilon}
\end{equation}
Hence, only close to an infrared fixed point situation, where $\Lambda \to \infty$ one has $\varepsilon \to 0^+$, and the duration of the scattering is much larger than all other scales in the problem.
In general, $1/\varepsilon$ must be kept finite, and this expresses the life time of the induced space time deformation, which we now proceed to discuss.

To this end, we take into account~\cite{kmw} that the anomalous dimension of the operators (\ref{recoilop}),(\ref{recoilopII}) is $-\varepsilon^2/2$ and hence these deformations are relevant from a world-sheet renormalization group viewpoint. The corresponding open string $\sigma$-model is thereby {\it non conformal}, in need of Liouville dressing~\cite{ddk} in order to restore the conformal symmetry and hence the consistency of the world-sheet theory~\cite{gsw}.  The dressed operator (\ref{recoilop}), which is the leading contribution to recoil in the limit of small $\varepsilon $, reads:
\begin{equation}
V^L_{\rm rec} = \int_{\partial \Sigma} e^{\alpha \varphi}u_i X^0 \Theta_\varepsilon (X^0) \partial^n X^i ~,
\label{dressedrecoil}
\end{equation}
where $\varphi $ is the Liouville mode~\cite{ddk}, and $\alpha  $ the Liouville anomalous dimension (\ref{lad}), which is such that the dressed operator (\ref{dressedrecoil}) is conformal (i.e. of conformal dimension (1,1)) on the world-sheet. Detailed analysis~\cite{emnrecoil} shows that there is a {\it central charge surplus} of order $\varepsilon ^4$, and hence the deformed $\sigma$-model is {\it supercritical}~\cite{aben} and the Liouville mode {\it time like}.

The Liouville anomalous dimension is, to leading order in $\varepsilon$ (c.f (\ref{solutions})): $\alpha \sim \varepsilon/\sqrt{2}$. By rewriting the boundary dressed operator (\ref{dressedrecoil}) as a bulk one, using Stokes theorem, and performing the relevant partial integrations on the world sheet, using also the (world-sheet) string equations of motion, we may arrive at the following bulk world-sheet $\Sigma$-deformation~\cite{emnrecoil}:
\begin{equation}
   V_{\rm bulk} \sim  -\varepsilon u_i \int_\Sigma e^{\epsilon \varphi/\sqrt{2}} X^0 \Theta_\varepsilon (X^0) \partial \varphi {\overline \partial} X^i \sim -\varepsilon u_i \int_\Sigma e^{\varepsilon (\varphi/\sqrt{2} - X^0)} X^0 \partial \varphi {\overline \partial} X^i
\label{dr2}
\end{equation}
where we approximated $\Theta_\varepsilon (X^0) \sim e^{-\varepsilon X^0}$, for $X^0 > 0$ where our formalism applies.
We now observe that for times of order  $ X^0 \sim \Delta t \sim 1/\varepsilon $, i.e. within the duration of the scattering, the bulk operator (\ref{dr2}) implies an off-diagonal deformation of the space-time metric~\cite{emnrecoil}, with $\varphi-i$ components
\begin{equation}
    g_{i\varphi } \sim e^{\varepsilon (\varphi/\sqrt{2} - X^0)} u_i \varepsilon X^0 ~, \qquad X^0 \sim \Delta t \sim  1/\varepsilon
\label{offdmetric}
\end{equation}
The Liouville dressing procedure increases the target space dimension by one, and in this particular case the string is supercritical, so the D+1 enhanced space time has in fact two time-like coordinates, $X^0$ (if the original space time had Minkowski signature) and $\varphi$. However, in the approach to recoil advocated in refs. \cite{kmw,emnrecoil}, for reasons of convergence of the world-sheet path integral, the coordinate $X^0$ is assumed of Euclidean signature. This implies the following form for the induced metric of the Liouville-augmented space-time:
\begin{equation}
   ds^2_{\rm Liouville~dress} = -(d\varphi )^2 + (dX^0)^2 + 2g_{i \varphi } d\varphi dX^i + (dX^i)^2
\label{dressmetric}
\end{equation}
In the approach of \cite{emn}, the ``physical time'' lies on an appropriate hyper-surface obtained by keeping fixed the quantity:
\begin{equation}
   \varphi + ({\rm const}) X^0 = 0
\end{equation}
where the constant in front of $X^0$ is positive and its value depends on the details on the underlying microscopic model.
The opposite flow of time with respect to the Liouville mode is a generic feature, related to specific renormalization group properties of the Liouville mode, which, in the approach of \cite{emn} is considered as a local world-sheet renormalization group scale. As such, it flows from an Infrared to ultraviolet and back to an infrared fixed point on the world sheet along specific closed-path  trajectories, dictated by the appropriate quantization procedure of the
Liouville theory~\cite{emn}. The corresponding target space flow then must be opposite, for reasons associated with the stability properties of the target-space effective action, as explained in detail in \cite{emn}, where we refer the interested reader for further reading.

In what follows we shall impose the constraint:
\begin{equation}
    \frac{\varphi}{\sqrt{2}} + X^0 = 0
\label{flow}
\end{equation}
The normalization in (\ref{flow}) is fixed by the requirement that after the identification (up to a positive proportionality constant) of the Liouville mode with minus the target time, the resulting D-dimensional space time
must have a  time coordinate normalized as in Robertson-Walker space times, which defines the physical time~\footnote{A more complicated situation, leading to different proportionality constants between Liouville mode and target time has been considered in \cite{horizon}, where a semi-microscopic model to describe D-particle recoil induced by strings propagating in brane worlds has been considered. The induced four-dimensional metric in such a situation is more complicated than (\ref{dressmetric}), as is obtained by averaging appropriately over statistical populations of D-particles (c.f. some discussion below). The results of both analyses, however, as far as  D-particle-recoil-induced modified dispersion relations for matter probes are  concerned, are qualitatively similar and, hence, for our purposes in this article we shall consider (\ref{flow}) from now on.}. Interestingly enough,  the combination on the left-hand-side of
(\ref{flow}) appears~\cite{gravanis} in the  mass squared terms of the effective potential of the low-energy theory in the four-dimensional space time obtained after compactifying the superstring model on magnetized torii, with a transient ``fictitious '' magnetic field in the extra dimensions, which asymptotes to an intensity $H$. Such transient fields might characterize collision of brane worlds in the way explained in \cite{gravanis}. It runs out that, if we parametrize the transient field strength by $H e^{\varepsilon X^0}$, for times $X^0 < 1/\varepsilon $, $\varepsilon \to 0^+$ (similar to our recoil case above) and constant afterwards, one arrives again at supercritical strings. After Liouville dressing, to restore conformal invariance, the mass-squared splittings $\delta m^2$ between fermionic and bosonic matter degrees of freedom  on the brane worlds, as a result of their coupling with the magnetic field $H$ (a sort of ``Zeeman effect''), are given by (for the example of toroidal compactification, say, of the 4th and 5th spatial directions of a five-brane world compactified to a three-brane~\cite{gravanis}):
\begin{equation}
    \delta m^2_{B-F} \propto He^{\varepsilon\frac{\varphi}{\sqrt{2}} + X^0}\Sigma_{45}
\label{split}
\end{equation}
with $\Sigma_{45}$ denoting spin structures in the plane of the compactification torus.
Stability of the theory, i.e. time independent masses, can be guaranteed, independently of the value of $\varepsilon $, if the constraint (\ref{flow}) is satisfied. In this way, the Liouville dressed theory, guarantees the asymptotic stability of the supersymmetry obstructing mass splittings beyond the time $1/\varepsilon$.

We may enforce the above scenario in our D-particle models, depicted in fig.~\ref{fig:recoil}, if we consider the three-brane world as being obtained from compactification of five (or higher-dimensional) branes on such magnetized manifolds, and assume that, at an early epoch, there was a collision with another brane. Such models can make consistent string backgrounds if the branes are still~\cite{emw}. The collision will also break target space supersymmetry, in a way determined by the magnitude of the fictitious magnetic field $H$, independently of the magnitude of the dark energy of the brane world~\cite{gravanis,emnw}, and hence desirable from a cosmological/phenomenological view point.

The recoil analysis, described above, goes in parallel with the supersymmetry obstruction scenario due to the magnetized compactification, and hence the
identification of time with the opposite of the Liouville mode (\ref{flow}) is guaranteed in a dynamical way, ensuring stability of the ground state of the system.

Upon this identification, the off-diagonal induced metric (\ref{offdmetric}) becomes off-diagonal in a four-dimensional space-time sense:
\begin{equation}
g_{0i} (X^0, u_i) \sim u_i \varepsilon X^0 e^{-2\varepsilon X^0}
\label{momentumtimemetric}
\end{equation}
For times of order of the duration $\Delta t$ of the scattering, $X^0 \sim \Delta t \sim 1/\varepsilon $, the order of magnitude of the metric is:
\begin{equation}
g_{0i} (X^0, u_i)\sim u_i ~, \qquad X^0 \sim \Delta t \sim 1/\varepsilon
\label{metricuncert}
\end{equation}
The metric, as expected, decays exponentially with the target time~\footnote{In the original works \cite{emnrecoil} we have considered times of order $1/\varepsilon$ so we ignored the decay, and concentrated only on the metric (\ref{metricuncert}). Here we keep the $X^0$-dependence explicit, in order to make manifest the relaxation nature of the induced deformation, implying an asymptotic Minkowski metric.} for $X^0 > 1/\varepsilon$, leading to an asymptotic Minkowski metric for times $X^0 \gg 1/\varepsilon $.

Taking into account that the intermediate-state(open-string/D-particle) recoil velocity $u_i$ is related to
a fraction of the incident momentum~\cite{kmw,emnrecoil,szabo}, $u_i = g_s k_i/M_s$, where $M_s/g_s$ is the mass of the D-particle, with $g_s$ the string coupling, and $M_s$ the string mass scale, we then observe that the induced metric (\ref{momentumtimemetric}) (or (\ref{metricuncert}) depends on {\it both} target space-time coordinates and momenta of the open string. In fact, immediately after the end of the uncertainty time period $\delta t$, where the open string state is re-emitted, with a momentum $k_{2,i}$, the D-particle recoils with a velocity $u_i' = k_{1,i} + k_{2,i}$, i.e the momentum transfer. Thus, the induced space time geometry also changes then. However, because the momentum transfer is of the order of the incident momentum, the induced metric deformation retains its order of magnitude. In the original works~\cite{emnrecoil}, and also in what follows, it is this geometry, after the re-emission of the string, that we consider explicitly.

The dependence of the metric (\ref{metricuncert}), (\ref{momentumtimemetric}) on both coordinates and momenta implies a Finsler construction~\cite{finslermetric}. It should be mentioned at this point that such a situation also characterizes~\cite{finsler} some Deformed Special Relativities (DSR)~\cite{dsr}, where the so-called
``rainbow metric''~\cite{rainbow} emerges as a momentum dependent metric to reproduce modified dispersion relations of a specific DSR model. We shall discuss further the comparison of our model with such approaches in the next section.

In general, the distorted space time geometry, as observed by an observer who is initially at rest with respect to the D-particle, during an uncertainty time interval $\Delta t \sim 1/\varepsilon $ describing the scattering period, will be given by (\ref{metricuncert}), where $u_i \sim r g_s k_i/M_s $, with $r < 1$, depending on the microscopic details of the momentum transfer during the
capture process of fig.~\ref{fig:recoil}.

The presence of a spatial recoil vector $u_i$ in the metric distortion (\ref{metricuncert}) implies {\it Lorentz symmetry breaking}, or, better {\it reduction.} In fact, as discussed in \cite{volkov} where we couple a fermionic system to this type of metrics, there is a {\it reduced symmetry} that characterizes the situation, namely a two-parameter subgroup of the $SL(2,C)$ symmetry that leaves the magnitude of the recoil velocity $\vec u$ invariant.

If we consider the situation~\cite{horizon} where many such D-particles puncture our brane world, as the latter sweeps through the bulk space time, then from the point of view of a three-dimensional brane observer, the D-particles will look like ``space-time foamy'' defect, flashing on and off with a frequency that depends on detailed dynamical properties of the complete bulk system.
It is for this reason that we call such models, with a statistical population of D-particles, ``D-particle foam''~\cite{emnrecoil,emnw}.
In such models, during the propagation of matter probes, there will be multiple scatterings of the probe with the D-particle defects. In each scattering there will be induced, {\it locally} in space-time, a coordinate- and momentum-dependent metric distortion of the form (\ref{metricuncert}).

As discussed in \cite{szabo}, the dynamics of the recoil degrees of freedom are described by an appropriate Abelian gauge field, which obeys a {\it subluminal} Born-Infeld type action. This latter feature implies that the recoil/capture  process depicted in fig.~\ref{fig:recoil} is {\it causal}, and there will be a delay in the probe's propagation, as compared to its propagation in a vacuum
without D-particles. The delay will occur as a result of the induced modified dispersion relation, for a probe of mass $m$, due to the metric (\ref{metricuncert}):
\begin{equation}
k_\mu k_\nu g^{\mu\nu}(k) = -m^2
\label{mdr}
\end{equation}
We should mention at this point that, in a statistical ensemble of $D$-particles, it is possible that the effects of the metric distortions due to recoil cancel out on average~\cite{sarben}, $\langle\langle u_i \rangle\rangle = 0$, but quantum fluctuations remain non-trivial, $\langle\langle u_i u_j \rangle\rangle = \sigma^2 \delta_{ij},~\sigma \ne 0$. In such a case the observable effects become much harder to detect.

A modified dispersion relation will lead, through the appropriate group velocity, to a non-trivial {\it vacuum refractive index} for massless probes propagating in the above D-particle foam space time. As discussed above, the
refractive index will be {\it sub-luminal}, due to special dynamics properties of the stringy recoil problem~\cite{szabo}. From the momentum dependence of the associated metric distortion (\ref{metricuncert}), we then observe that, unless the effects cancel out upon averaging over D-particle populations~\footnote{
If, of course, the D-particle foam effects cancel on average as far as momentum transfer is concerned during the multiple scatterings,  $\langle\langle u_i \rangle\rangle =0$, then the quantum fluctuations may induce much smaller in magnitude terms in the Quantum Gravity modifications to the dispersion relation, suppressed by quadratic or higher powers of the string scale. As mentioned above, such terms are much harder to detect by present or immediate-future facilities, although one cannot exclude the possibility of reaching such sensitivities in the future if one detects and uses as probes very energetic cosmic particles, such as ultra-high-energy neutrinos from gamma ray bursts {\it etc.}~\cite{poland}.},
we are faced with a refractive index of the form (\ref{refindex}), exhibiting minimal (linear) suppression by the string scale:
\begin{equation}
           v_{\rm massless~probe-D-recoil} = 1 - \xi g_s\frac{|\vec k|}{M_s}
\label{refindexrecoil}
\end{equation}
where the parameter $\xi > 0$ depends on the details of the D-particle foam, such as the D-particle density~\cite{horizon}.
Such details cannot be estimated by theoretical considerations at present, as we are lacking a fundamental microscopic model.
However, progress towards this direction is made by constructing explicitly superstring/super-brane world models in higher dimensional bulk space times~\cite{emnw}.
The big issue pending of course, towards phenomenologically realistic four dimensional models, is the appropriate compactification procedure, which unfortunately entails all sorts of complications regarding supersymmetry breaking {\it etc.}

The main conclusion from this section so far, therefore, is that in the non-critical stringy models of space-time foam, involving D-particles, there is an induced
modification to the space time metric,  (\ref{metricuncert}), which depends {\it linearly} on the incident momentum, and eventually leads to {\it subluminal} modified dispersion relations (\ref{mdr}) on average. This subluminality feature will distinguish the approach from others in LIV Quantum Gravity,
such as certain loop quantum gravity models and other proposals~\cite{alfaro}, where both subluminal and superluminal propagation may be allowed, thereby implying experimentally detectable (in principle) birefringence effects.

It has been observed~\cite{ems}, though, that not all string states, may be captured by D-particles. In particular, electrically charged excitations
do not exhibit that behaviour, as a result of conservation of the electric flux and charge. In this respect, electrons might not exhibit such a behaviour in the presence of D-particle foam, which might help avoiding the stringent constraints on
linearly suppressed (with the QG scale) dispersion relations, coming from studies of synchrotron radiation from Crab Nebula~\cite{crab}. The latter, however, cannot place any stringent constraints on photon dispersion.
In our approach, photons, as electrically neutral, can interact with the D-particles in the topologically non-trivial way depicted in fig.~\ref{fig:recoil}, involving splitting of strings, and, as such, they  constitute one of the most sensitive probes of such modified dispersion relations. Of particular use are tests in which one measures arrival times of photons with different energies from gamma ray bursts, assuming simultaneous emission of
photons with various energies
at the source~\cite{grb}. According to the subluminal dispersion relation (\ref{mdr}), the highly energetic photons will be {\it delayed } more, as compared with the less-energetic ones.

We also remark that the D-particle recoil model can be used as a model for the implementation of relativistic modified Newtonian dynamics in a string theory context~\cite{sakellariadou}. Indeed, if we assume that there is no uniform distribution of D-particles in the brane universe, but that the latter, as being massive with masses $M_s/g_s$, are concentrated more near galaxies, then due to the recoil process, one would obtain
an induced modification to the regional metric, which,
to lowest order in the recoil velocity, would be given by (\ref{metricuncert}). In \cite{sakellariadou} we have extended the analysis to discuss higher order corrections to the recoil-induced metric, which we rewrote in a covariant form. We came to the conclusion that world-sheet conformal invariance can be maintained for metrics of the form:
\begin{equation}
   g_{\mu\nu}^{\rm matter} = e^{-2\Phi} g_{\mu\nu} + f(\Phi) (u_\mu u_\nu)
\label{bimetric}
\end{equation}
where $g_{\mu\nu}$ is an Einstein-frame~\cite{aben} metric,
$u_\mu$ is the four-vector of the D-particle recoil velocity, satisfying the constraint $u_\mu u_\nu g^{\mu\nu}= -1$, and
$f(\Phi )$ is an appropriate function of the dilaton $\Phi$, which can be determined by demanding smooth connection of the metric
$g_{\mu\nu}^{\rm matter} $ with the Einstein metric $g_{\mu\nu}$, both,
at the moment of impact (fig.~\ref{fig:recoil})  and asymptotically in target time.
In \cite{sakellariadou} we discussed explicitly the case of linear dilatons, $\Phi \sim u_\mu X^\mu$, and flat Minkowski metrics $g_{\mu\nu} = \eta_{\mu\nu}$, but this should be considered only as a toy situation. The full problem requires complicated, time dependent dilaton backgrounds, which should satisfy world-sheet conformal invariance conditions together with the appropriate Einstein graviton backgrounds $g_{\mu\nu}$. This general problem is still open.

The important point of (\ref{bimetric}) is that it implies a bi-metric theory, since, as mentioned previously, not all string states interact non trivially with the D-particle foam in the way demonstrated in fig.~\ref{fig:recoil}.
The induced metric (\ref{bimetric}) would lead to modified Newtonian  dynamics in the regions where the concentration of the D-particle foam would be largest. In \cite{sakellariadou} we assumed that such regions are near galaxies, thereby leading to modified newtonian dynamics at galactic scales, in a similar spirit to the ideas advocated in \cite{bekenstein}, where metric deformations of the form (\ref{bimetric}) have been proposed, but with the r\^ole of $u_\mu$ played by arbitrary Lorentz violating vector fields $A_\mu$.

The dynamics of the $u_\mu$ recoil four vectors, though, in string theory is different from the one of the $A_\mu$-field in \cite{bekenstein}. Indeed, as already mentioned, the recoil degrees of freedom obey a Born-Infeld non-linear Lagrangian~\cite{szabo}. The model has important implications for the amount of dark matter required in the Universe. In fact, by assuming a situation in which, on average over large (statistically significant) populations of D-particles in galaxies, there is only a temporal time-dependent (isotropic) component of $u_\mu$, we came to the conclusion~\cite{sakellariadou} that a substantial amount of neutrino dark matter should be present. This is in contrast to the models in \cite{bekenstein}, which used the modified Newtonian dynamics at galactic scales to reproduce the relevant rotational curves and thus argue in favour of the absence of dark matter altogether. As explained above, neutrinos are among the few neutral particles that interact non trivially (c.f. fig.~\ref{fig:recoil}) with the D-particle foam, and as such they induce recoil, affecting significantly the galactic dynamics~\footnote{Moreover, since neutrinos avoid clustering, they have also been conjectured in \cite{sakellariadou} to contribute to the dark energy of the Universe in a non-trivial way, which I do not have the time or space to explain here.}. Our analysis is closer in spirit to that of ref.~\cite{skordis}, where, in order to reproduce the acoustic peaks in the CMB spectrum, the authors have also included neutrino dark matter in the models of \cite{bekenstein}.
For further details on this scenario we refer the reader to \cite{sakellariadou}, as further discussion on such issues would take us out of the main scope of the present article.

A final but important comment we would like to make before closing this section concerns neutrinos or in general electrically neutral ``flavoured'' particles. Here, by the word flavour we do not necessarily mean neutrino flavour, but we use it also to describe neutral meson CP eigenstates (different from mass eigenstates), e.g. for kaons  there are long and short-lived CP eigenstates, $K_L^0$, $K_S^0$ {\it etc.} If the scattered string state in fig.~\ref{fig:recoil} represents such a flavoured state (even a composite one, like a neutral kaon), then the re-emitted open string state after the recoil might be of a different flavour, given that flavour may not be conserved during the capture/recoil process by the D-particle~\cite{sarben}. Although to prove this rigorously one needs a detailed microscopic description of such excitations in the context of the relevant string model, which is still lacking, nevertheless it looks a reasonable assumption to make, following other similar treatments of space-time foamy situations involving virtual black holes, which also do not conserve flavour~\cite{poland}.

This proves important for inducing possibly unique effects on entangled states of particles. Indeed, as discussed in detail in \cite{bms}, this flavour non-conservation in D-particle foam models results in a particular violation of CPT symmetry, associated with the
fact that the generator of CPT transformations is not well defined, at least in the framework of effective theories. This induces modifications to the pertinent Einstein-Podolsky-Rosen (EPR) correlations of particles~\cite{bmp,poland}. To understand briefly what happens, consider the case of a $\phi$ meson decay in a so-called $\phi$-factory. If normal CPT symmetry is assumed, the decay produces EPR entangled states of $K_L K_S$ on both sides of the detector. However, in the D-particle foam model of \cite{bms},
characterized by the above-mentioned specific type of CPT violation,
the results are contaminated by $K_S K_S$ and $K_L K_L$ pairs. In fact, one can distinguish such genuine quantum gravity effects from background effects, due to their properties~\cite{bmp}. Hence, if observed, such effects would constitute, in my opinion, ``smoking-gun'' evidence of this type of CPT violation.

\section{Comparison with strings in Non-Commutative space times}

Lorentz violation may characterize non-commutative space times, as is for instance the case of space times with background electric and/or magnetic fields. Consider, for instance, the generic non commutativity relation among the coordinates $x^\mu$ of a manifold:
\begin{equation}
     [x^\mu, x^\nu] = i\theta^{\mu\nu}
\label{ncom}
\end{equation}
with $\theta^{\mu\nu}$ real and antisymmetric in its indices.The presence of $\theta$ imply generically violations of Lorentz symmetry. From the point of view of strings, which is of interest to us here, we mention that non commutative (Lorentz violating) space times do arise in many instances in string theory, notably in cases with non trivial background electric or magnetic fields, with fixed direction in space.

It has been argued~\cite{noncom} that any realistic non commutative field theory, endowed with its proper Moyal $\star$ product ($~f \star g (x) \equiv {\rm exp}\left(\frac{i}{2}\theta^{\mu\nu}\partial_{x^\mu}\partial_{x^\nu}\right)f(x)g(y)|_{x=y}~$), can be made physically equivalent to a subset of a general Lorentz violating extension of the standard model (SME), of the type considered by Kostelecky and collaborators~\cite{sme} and mentioned in section 2 of the present article.
It is interesting to notice~\cite{noncom} that, since many non commutative field theories satisfy CPT invariance, the resulting  SME effective theories involve non renormalizable Lorentz violating terms, which however respect CPT symmetry.
For example, regarding SME quantum electrodynamics, which was the subject of our discussion in section 2, the following subset of terms may be physically equivalent to a CPT conserving non-commutative field theory to leading order in $\theta^{\mu\nu}$~\cite{noncom}:
\begin{eqnarray}
\cL &=&
\half i \overline{\ps} \ga^\mu \lrDmu \ps
- m \overline{\ps} \ps
- \frac 1 4 F_{\mu\nu} F^{\mu\nu}
- \frac 1 8 i q\th^{\al\be} F_{\al\be}
\overline{\ps} \ga^\mu \lrDmu \ps
+ \frac 1 4 i q\th^{\al\be} F_{\al\mu}
\overline{\ps} \ga^\mu \lrDbe \ps
\nonumber\\
&&
+ \frac 1 4 m q \th^{\al\be}F_{\al\be} \overline{\ps} \ps
- \frac 1 2 q \th^{\al\be} F_{\al\mu} F_{\be\nu} F^{\mu\nu}
+ \frac 1 8 q \th^{\al\be} F_{\al\be} F_{\mu\nu} F^{\mu\nu}.
\label{ncqed}
\end{eqnarray}
with $q$ being the electric charge and $D_\mu $ denoting the gauge covariant derivative as usual. The reader should compare (\ref{ncqed})
with the corresponding expression for SME, leading to (\ref{diracsme}): all the CPT-violating terms are absent, since the microscopic non commutative
field theory is CPT invariant.

For our purposes in this work, we mention that in general, non-commutative field theories appear to have problems with either unitarity and/or causality~\cite{toumbas}. Indeed, the authors of \cite{toumbas} considered
the elastic scattering of two wavepackets into two outgoing ones, in a generic non-commutative field theory (\ref{ncom}), and demonstrated the existence of acausal (``advanced'') out-going wave packets, with negative delays $\Delta t$, i.e. occurring before even the scattering of the incident waves took place. Also their calculation showed that rigid rods grow instead of Lorentz contracting at high energies.

This unphysical situation, which seems to be generic to non commutative field theories, is eliminated if one considers stringy effects. Indeed, the authors of \cite{toumbas} repeated the scattering calculation by considering wavepackets in open (super) string theory. The result shows no advanced waves, and more interestingly, the outgoing wave packet splits into a series of packets, one located at the origin, $x =0$, and the others at $x = 16\pi n \alpha ' p_0$ (where $p_0$ is the energy of the wave packet, $\alpha '=\ell_s^2$ is the Regge slope, and $\ell_s$ is the string length scale),
with increasing spread and decreasing amplitude as the order $n$ of the packet increases.  There is an intermediate stretched string state formed, which oscillates, thereby producing the series of the outgoing packets. The string has total energy $p_0$, and its length begins to grow up to order $L \sim \alpha ' p_0$  storing the energy as potential energy. However, the scattering is {\it causal}, and the positive time delay, obtained in the calculation, increases linearly with the energy $p_0$:
\begin{equation}
\Delta t = \alpha ' p_0~,
\label{wpuncert}
\end{equation}
which is consistent with the string uncertainty principle~\cite{yoneya}
\begin{equation}
      \Delta L \Delta t \ge \alpha '~.
      \label{stringuncert}
      \end{equation}
It is the above space-time uncertainty that makes a generic qualitative connection of open strings with non commutative space time theories. This connection is quantified when we consider the strings in background electric fields $|\vec E |$ for instance. In such a case, the above-mentioned scattering is also found causal, with the only modification~\cite{toumbas} that the time delay (\ref{wpuncert}) is multiplied  by a factor $1/(1 - |\vec E |^2)$.

Moreover, for high energies, higher-genus (quantum string corrections) amplitudes should be considered~\cite{genus}, since these are the dominant ones. If one repeats the calculation for the four-point open-string amplitude on a fixed genus $G$ world-sheet, then, the relevant causal time delays are reduced~\cite{genus}: $\Delta t \sim \frac{\alpha ' p_0}{G+1}$. We mention at this point that Genus re-summation effects are still not fully understood in string theory (due to issues regarding re-summability of the genus expansion), and hence truly non perturbative expressions for the above time delays are not known at present.

In tests of photon dispersion relations, such as those in \cite{grb} studying the arrival times of photons from distant gamma-ray bursts, uncertainties due to critical-string interactions, as in (\ref{wpuncert}), should be considered as source effects, describing for instance interactions among photons - viewed as open string states- at the source regions. Such uncertainties would result in non simultaneous emissions of photons, and should be taken in principle into account
in studies like \cite{grb}, in order to correct the analyses in searches for quantum-gravity propagation effects. However, even for large string length scales, $\ell_s \gg \ell_{\rm Planck}$, the number of total photon-photon scatterings at the source are such that these effects are negligible, when compared to the currently available time resolutions in these measurements.

The time delays/uncertainties (\ref{wpuncert}), that increase {\it linearly} with the energy, bear some formal similarities with our D-particle/open-string-state interactions. There again, we have had the formation of an intermediate composite state (see fig.~\ref{fig:recoil}), but the difference from the flat-space-time case of open-string scattering, was the induced Finsler-like metric (\ref{metricuncert}). Nevertheless, precisely due to such metrics, there will be time delays for highly energetic photons, as compared to the less energetic ones, which will grow linearly with the energy, as follows from the induced refractive index effects (\ref{refindexrecoil}).
We note that, in the D-particle case, the total time delay of a matter probe in the D-particle foam, depends on the details of the defects distribution. If the latter is non uniform, as in the model of \cite{sakellariadou}, where the D-particle concentration is high near massive celestial bodies, due to the D-particle mass $M_s/g_s$, and very low in the ``empty'' space outside the body, then the arrival time delay effects due to (\ref{refindexrecoil}) are suppressed in GRB photon tests~\cite{grb}, compared to the case of uniform D-particle distribution over the D-brane world. This is a consequence of the fact that, in such a case, the effects are largest near the massive source, whilst photon propagation between the source and the detector is virtually free. However, it seems to me that, to obtain such source-D-foam induced delays that could be measurable by the current technology, would require an unphysically large concentration of D-particles near the source of the GRB.

\section{Brief Comparison with Deformed Special Relativities and models with Reduced Lorentz Symmetry: Finsler geometry as a common link}

Before closing the discussion, I would like to make a brief comparison of the above-mentioned results on momentum-dependent induced metrics in non-critical strings with a similar situation encountered in Deformed Special Relativities~\cite{dsr}.
I will be very brief, as such theories have been discussed at length in the conference by other speakers and also in the literature, where I refer the reader for details.

In ref. \cite{rainbow} it was pointed out that the dispersion relations from a specific form of doubly special relativity with an invariant energy scale can be represented as a dispersion relation on a ``rainbow'' metric, that is a metric which depended on both coordinates and momenta. As we have seen above, this was precisely the situation characterizing our recoil-induced metrics~\cite{emnrecoil} (\ref{metricuncert}), (\ref{bimetric}). The finslerian-geometry~\cite{finslermetric} interpretation given to such metrics in \cite{finsler} applies intact to our case as well.
In this respect, our D-particle recoil example may be considered as a case of
Finsler geometry in string theory, a topic which has been discussed in the context of heterotic strings in the past~\cite{vacaru}, but from a different perspective.

Finsler geometry is a generalization of the Riemannian geometry, based on the definition of a norm $F(x, u)$ which is a function of coordinates and also of a tangent vector $u$, and which replaces the usual structure of the inner product over the tangent bundle in the Riemannian geometry.
The norm $F(x, u)$ defines the metric~\cite{finslermetric,finsler}:
\begin{equation}\label{norm}
g_{\mu\nu} = \frac{1}{2}\frac{\partial^2 F^2}{\partial u^\mu \partial u^\nu}
\end{equation}
which can be equivalently expressed via the relation:
\begin{equation}
F(x, u) = \sqrt{g_{\mu\nu}(x,u) u^\mu u^\nu}
\label{norm2}
\end{equation}

A particle, of mass $m$, moving in a Finsler geometry is characterized by an action
\begin{equation}\label{finaction}
I = m\int_{\rm i}^{\rm f} F (x, {\dot x}) d\tau
\end{equation}
where the overdot denotes derivative with respect to time $t$, and $F$ is the Finsler norm defined above (\ref{norm}), (\ref{norm2}). The norm can depend on several physically important parameters, such as the mass of the particle, or, as is the case with the D-particle recoil metric, the string scale and coupling, and in general the Planck scale for QG-induced modified dispersion relations.

The canonical momenta in this formalism are defined as~\cite{finsler}:
\begin{equation}\label{finmom}
  p_\mu = m\frac{\partial F}{\partial {\dot x}^\mu}=m\frac{g_{\mu\nu}(x,{\dot x}){\dot x}^\nu}{F}
\end{equation}
and the modified dispersion relations are obtained as a result of the mass-shell condition with respect to the canonical momenta
\begin{equation}\label{finmdr}
  h^{\mu\nu}(x,p) p_\mu p_\nu = -m^2
  \end{equation}
where $h^{\mu\nu}(x,p)$ is the inverse of the velocity-dependent metric,
$h^{\mu\nu} (x,p) = g^{\mu\nu}(x, {\dot x})$.

One can define appropriately geodesics in such space times, and discuss the associated symmetries. For details I refer the reader to the literature~\cite{finslermetric,finsler}. I only mention here that, the reduced Lorentz symmetry, for instance, of the recoil problem, that leaves the amplitude of the D-particle recoil-velocity vector
$|\vec u |$ invariant~\cite{volkov}, finds a natural interpretation in terms of the appropriate Finsler-metric symmetries, describing the modified dispersion relations of the D-particle recoil problem.

The above considerations were classical. However,
one might consider a semi-classical treatment of the induced Finsler geometries in non-critical strings by considering~\cite{emnrecoil,horizon} quantum fluctuations of the recoil velocity $u_i$. In first-quantized string theory, such fluctuations are obtained by summing up world-sheet topologies of higher genus. It may be the case that, on average, Lorentz symmetry is restored, $< u_i >=0$, but quantum fluctuations do not respect the symmetry, $<u_i u_j> \ne 0$ . In this spirit, we also encountered in the previous section a situation involving a statistical population of D-particles, in which case again Lorentz symmetry could be restored statistically, with its breaking showing up only in higher-order correlators.
Similar issues, regarding fluctuating geometries and non-trivial modified dispersion relations, have also been considered in a generic context in \cite{grillo}.

I would like now to discuss a final case of Lorentz violation,
where again Finsler geometry makes its appearance. The situation concerns
the so-called ``very special relativity'' (VSR) model of Cohen and Glashow~\cite{cohen}. VSR models are characterized by a reduced Lorentz symmetry, which is determined as follows: defining light-cone coordinates
$x^\pm = x^0 \pm x^3$ and $x^i$ for $i=1,2$ in a four-dimensional space-time,
the local physics symmetry group is provided by the following SIM(2) {\it subset} of Lorentz generators ${\cal M}_{\mu\nu}$:
\begin{equation}
                       \{ {\cal M}_{+-},~{\cal M}_{+i},~{\cal M}_{12}\}
\label{subset}
\end{equation}
i.e omitting ${\cal M}_{-i}$. Upon taking the direct product with the translations $\{P_+, P_-, P_i \}$,  the resulting subgroup, called ISIM(2),
leaves invariant the {\it null} direction
\begin{equation}
                      \eta^\mu = \delta^\mu_+
\label{null}
\end{equation}
in the sense that, under the action of ${\cal M}_{+-}$, $\eta^\mu $ is rescaled to $\lambda \eta^\mu $, for $\lambda \in R$.

In this picture there is an ``aether'' but moving at the speed of light, without definite velocity. In this sense it would be hard to detect experimentally.

In \cite{pope}, the authors looked for appropriate deformations of ISIM(2) in order to generalize the algebra to a curved background, and thus connect the approach to gravity. The proposed deformations, DISIM$_b$(2), are a subgroup of the Weyl group, involving the Poincare group and Dilations, and are represented by the action of ${\cal M}_{+-}$ as follows:
\begin{equation}
          x^i \rightarrow \lambda^{-b} x^i~, \qquad x^\pm \rightarrow \lambda^{-b\mp 1}x^\pm
          \end{equation}
with $b$ real, the parameter of the deformation.

It was observed~\cite{pope}, then, that the above construction was related not to gravity but to a Finslerian geometry of the type proposed in \cite{bogosl}, which leave invariant the following element:
\begin{equation}
         ds = \left(\eta_{\mu\nu}dx^\mu dx^\nu \right)^{(1-b)/2} \left(\eta_\rho dx^\rho \right)^b~\, \qquad \eta^\rho = \delta^\rho_+
\label{findisim}
\end{equation}
The authors of \cite{pope}, went further to construct a point-like particle action invariant under DISIM$_b$(2), following the generic procedure, described above (\ref{finaction}),  of constructing actions using Finsler norms (\ref{norm2}):
\begin{equation}
     I_{{\rm DISIM}_b{\rm (2)}} = -m \int dt \left(-\eta_{\mu\nu}{\dot x}^\mu {\dot x}^\nu \right)^{(1-b)/2}\left(-\eta_\rho {\dot x}^\rho \right)^b
\label{disimaction}
\end{equation}
Constructing the canonical momenta, as in (\ref{finmom}), then, yields
the mass-shell condition (\ref{finmdr}) for this case as:
\begin{equation}
                 \eta^{\mu\nu} p_\mu p_\nu = -m^2(1-b^2)\left(-\frac{\eta^\nu p_\nu}{m(1-b)}\right)^{2b/(1 +b)}
\end{equation}
implying modified dispersion relations, as generically expected in a Finsler geometry.

Bounds on $b$ can be found by looking for anisotropy limits in mass kinetic terms $\frac{1}{2}m_{ij} {\dot x}^i {\dot x^j} $ obtained from (\ref{disimaction}) by expanding in powers of (small) $b$. The current experimental situation yields the following bounds $\delta m/m \le 10^{-23}$, which imply very small deformation parameters~\cite{pope}: $|b| < 10^{-23}$, thereby calling for microscopic explanations (the problem is similar in spirit to the need for an explanation of the smallness of the cosmological constant).

The reader's attention is now called to a comparison of the above results for VSR with our reduced-Lorentz symmetry situation encountered in D-particle recoil~\cite{emnrecoil}, described previously. In the heavy (non-relativistic) D-particle foam case , as we have seen above, we have encountered a situation with reduced Lorentz symmetry for matter propagation~\cite{volkov}, but, there, the symmetry left invariant the magnitude of the recoil velocity 3-vector $\vec u $. However, when one considers statistically significant populations of D-particles, we have seen that there is the possibility of not having a definite spatial direction, but on average $\langle\langle u_i \rangle\rangle = 0 $ and $\langle\langle u_i u_j \rangle\rangle \propto \sigma^2 \delta_{ij}~, \sigma  \ne 0$ (the same might happen due to quantum fluctuations, which could wash out a spatial $u$-component average~\cite{emnrecoil,horizon}).
In the covariant generalization (\ref{bimetric}) of the recoil-induced geometry, advocated in \cite{sakellariadou}, we have seen that in such situations one could have only a temporal direction of the four vector $u_\mu$, which does not destroy the universe's isotropy and is left invariant by the symmetry transformations. Alternatively, one might consider adaptations of the model to situations in which one of the light-come directions (like the forward one $\delta^\mu_+$, as in the VSR case above) is fixed. In such a case, it would be interesting to investigate the nature of the appropriate (deformed) symmetry group, along the lines of the study for the VSR model~\cite{pope} .

I note that in the D-particle recoil case, as in VSR, one obtains deformations from a flat Minkowski metric, and in this sense only Finslerian geometries rather than full gravity emerge. However, I doubt that there is a no go theorem, and I firmly believe that the above Finslerian deformations do play an important r\^ole on the local symmetry group of quantum gravity, which might be a deformed symmetry like the cases encountered in DSR models.
As far as strings are concerned, our D-particle foam case is only a special example~\footnote{I note here that an attempt to discuss deformed Lorentz symmetries of the type appearing in DSR models~\cite{dsr}, leading to modified dispersion relations, in world-sheet $\sigma$-models, was also made in \cite{smolstrings}.}. In general, as mentioned previously, Finsler geometry is known to play a r\^ole in heterotic strings~\cite{vacaru}, and it would be interesting to discuss generalizations of the analysis of \cite{pope} in order to construct a first-quantized stringy world-sheet action, extending appropriately the particle-case result (\ref{disimaction}). I conjecture that Finsler geometry plays a fundamental r\^ole in the dynamics of M theory constructions.

\section{Conclusions and Outlook}

In this talk I discussed some instances in string theory where Lorentz Invariance Violations (LIV) may occur. I have discussed several topics which might be of interest to this conference. Namely, I first discussed LIV in the context of non-supersymmetric open string field theory, whose low energy limit is described by an extension of the standard model. Then I proceeded to a discussion of induced modifications in the dispersion relations for matter probes in the context of non-critical string theory and, as a specific example, I described D-particle foam models, in which some string states can be captured by point-like D-brane defects in space time. This procedure entails the distortion of space time in the neighborhood of the recoiling defect by a coordinate and momenta dependent Finsler-type metric, responsible for modified dispersion relations of low-energy string matter propagating in the foam. In fact the
induced metric may be such that one obtains non-trivial subluminal refractive indices, with corrections that depend linearly on the probe's energy, and are suppressed by a single power of the string scale.

I compared the situation with strings in non commutative space times, where causal scattering of string states also induce time uncertainties that grow linearly with the energy of the probe, as in the D-particle recoil case.
I also noticed that Finsler metrics appear in DSR, and also in some very special relativities, advocated recently, which preserve most of the
phenomenology of Lorentz invariant theories, but are characterized by a reduced symmetry.

I made the point that, although the phenomenology of LIV is complex, and there is no single figure of merit to parametrize such violations, nevertheless the key to obtain ``smoking-gun evidence'' of specific models might lie on the violation of quantum discrete symmetries, such as CPT. I mentioned briefly a specific way of breaking CPT symmetry, characterizing D-particle foam models, in which the CPT generator is not well defined as a quantum mechanical operator, and as such, it leads to unique effects on the entangled states of neutral mesons.

Although it is too early to speculate on the ability of detecting possible effects of quantum gravity in the near future, I believe that the current theoretical and experimental situation appears exciting, and surprises may indeed be waiting ``around the corner''. Many branches of modern physics, from astrophysics and particle physics (including strings, general relativity and cosmology in its theoretical subjects), to delicate atomic physics precision experiments, are currently participating in the quest for this elusive theory.
This multidisciplinary aspect has been captured well by the current conference, which brought together experimenters and theorists from many of the above disciplines to draft up future strategies for research and development in Quantum Gravity. Let us hope that their efforts will be rewarded by success in the not-so-distant future.


\begin{thebibliography}{99}



\bibitem{coleman}
  S.~R.~Coleman and S.~L.~Glashow,
  Phys.\ Rev.\  D {\bf 59} (1999) 116008
  [arXiv:hep-ph/9812418].


\bibitem{kosstring} V.~A.~Kostelecky and S.~Samuel,
  Nucl.\ Phys.\  B {\bf 336} (1990) 263;
  Phys.\ Rev.\  D {\bf 39} (1989) 683;
  V.~A.~Kostelecky and R.~Potting,
  Nucl.\ Phys.\  B {\bf 359} (1991) 545;
  Phys.\ Lett.\  B {\bf 381} (1996) 89.

\bibitem{emn} J.~R.~Ellis, N.~E.~Mavromatos and D.~V.~Nanopoulos,
Phys.\ Lett.\ B {\bf 293}, 37 (1992)
[arXiv:hep-th/9207103];
{\it Invited review for the special Issue of J.\ Chaos Solitons Fractals},
Vol.\ 10, (eds. C. Castro amd M.S. El Naschie,
Elsevier Science, Pergamon 1999) 345
[arXiv:hep-th/9805120];
Phys.\ Rev.\ D {\bf 63}, 024024 (2001)
[arXiv:gr-qc/0007044].


\bibitem{emnrecoil} J.~R.~Ellis, N.~E.~Mavromatos and D.~V.~Nanopoulos,
  Phys.\ Rev.\  D {\bf 61} (2000) 027503
  [arXiv:gr-qc/9906029];
  Int.\ J.\ Mod.\ Phys.\  A {\bf 13} (1998) 1059
  [arXiv:hep-th/9609238].

\bibitem{dsr} G.~Amelino-Camelia,
  Int.\ J.\ Mod.\ Phys.\  D {\bf 11} (2002) 35
  [arXiv:gr-qc/0012051];
  Nature {\bf 418} (2002) 34
  [arXiv:gr-qc/0207049];
J.~Magueijo and L.~Smolin,
  Phys.\ Rev.\ Lett.\  {\bf 88} (2002) 190403
  [arXiv:hep-th/0112090];
  Phys.\ Rev.\  D {\bf 67} (2003) 044017
  [arXiv:gr-qc/0207085]. For reviews
  see:  G.~Amelino-Camelia and J.~Kowalski-Glikman (eds.),
  {\it Planck Scale Effects In Astrophysics And Cosmology}, Proc. 40th
  Karpacs Winter School, Ladek Zdroj, Poland, February 4-14, 2004 (Springer 2005);
also talk by J. Kowalski-Glikman, these proceedings and references therein.

\bibitem{finslermetric} see for instance: D. Bao, S.S. Chern and Z. Shen, {\it An introduction to Riemann-Finsler Geometry}, Springer-Verlag (NY, 2000).

\bibitem{finsler} F.~Girelli, S.~Liberati and L.~Sindoni,
  Phys.\ Rev.\  D {\bf 75} (2007) 064015
  [arXiv:gr-qc/0611024].


\bibitem{jacobson}
  T.~Jacobson, S.~Liberati and D.~Mattingly,
  Annals Phys.\  {\bf 321} (2006) 150
  [arXiv:astro-ph/0505267] and references therein;
see also: S.~Liberati,
  PoS {\bf P2GC} (2007) 018
  [arXiv:0706.0142 [gr-qc]];
  C.~Eling, T.~Jacobson and D.~Mattingly,
  arXiv:gr-qc/0410001.

\bibitem{poland} N.E.~Mavromatos,
  Lect.\ Notes Phys.\  {\bf 669}(2005) 245
  [gr-qc/0407005] and references therein;
  arXiv:0707.3422 [hep-ph], to appear in Proc. of Science (Kaon07 international conference, Frascati (Italy), May 21-25 2007);
J.~Bernabeu, J.~Ellis, N.~E.~Mavromatos, D.~V.~Nanopoulos and J.~Papavassiliou,
  arXiv:hep-ph/0607322, \emph{III Da$\Phi$NE Physics Handbook}, in  press.


\bibitem{sme} V.A.Kostelecky,
{\it CPT and Lorentz Symmetry III} (World Scientific, Singapore
2005) and references therein;
  Phys.\ Rev.\ Lett.\  {\bf 80} (1998) 1818.


\bibitem{pope}
  G.~W.~Gibbons, J.~Gomis and C.~N.~Pope,
  arXiv:0707.2174 [hep-th].

\bibitem{cohen}
  A.~G.~Cohen and S.~L.~Glashow,
  Phys.\ Rev.\ Lett.\  {\bf 97} (2006) 021601
  [arXiv:hep-ph/0601236].



\bibitem{bms} J.~Bernabeu, N.~E.~Mavromatos and Sarben~Sarkar,
  Phys.\ Rev.\  D {\bf 74} (2006) 045014.


\bibitem{gsw} M.B. Green, J.H. Schwarz and E. Witten, {\it Superstring
Theory}, Vols. I \& II (Cambridge University Press, 1987).


\bibitem{ddk} F.~David,
Mod.\ Phys.\ Lett.\ A {\bf 3}, 1651 (1988);
J.~Distler and H.~Kawai,
Nucl.\ Phys.\ B {\bf 321}, 509 (1989);
J.~Distler, Z.~Hlousek and H.~Kawai,
Int.\ J.\ Mod.\ Phys.\ A {\bf 5}, 391 (1990);
see also: N.~E.~Mavromatos and J.~L.~Miramontes,
Mod.\ Phys.\ Lett.\ A {\bf 4}, 1847 (1989);
E.~D'Hoker and P.~S.~Kurzepa,
Mod.\ Phys.\ Lett.\ A {\bf 5}, 1411 (1990).


\bibitem{osborn} H.~Osborn,
  Phys.\ Lett.\  B {\bf 222}, 97 (1989).

\bibitem{zam} A.~B.~Zamolodchikov,
JETP Lett.\  {\bf 43}, 730 (1986)
[Pisma Zh.\ Eksp.\ Teor.\ Fiz.\  {\bf 43}, 565 (1986)].


\bibitem{emninfl} J.~R.~Ellis, N.~E.~Mavromatos and D.~V.~Nanopoulos,
Mod.\ Phys.\ Lett.\ A {\bf 10}, 1685 (1995)
[arXiv:hep-th/9503162];

\bibitem{emnw} J.~R.~Ellis, N.~E.~Mavromatos, D.~V.~Nanopoulos and M.~Westmuckett,
  Int.\ J.\ Mod.\ Phys.\  A {\bf 21} (2006) 1379
  [arXiv:gr-qc/0508105] and references therein.


\bibitem{aben} I.~Antoniadis, C.~Bachas, J.~R.~Ellis and D.~V.~Nanopoulos,
Phys.\ Lett.\ B {\bf 211}, 393 (1988);
Nucl.\ Phys.\ B {\bf 328}, 117 (1989);
Phys.\ Lett.\ B {\bf 257}, 278 (1991).

\bibitem{recentssc} O.~Aharony and E.~Silverstein,
  Phys.\ Rev.\  D {\bf 75} (2007) 046003
  [arXiv:hep-th/0612031];
  S.~Hellerman and I.~Swanson,
  arXiv:hep-th/0612116;
  E.~Tetteh-Lartey,
  Phys.\ Rev.\  D {\bf 75} (2007) 106005
  [arXiv:hep-th/0703160];
  G.~A.~Diamandis, B.~C.~Georgalas, A.~B.~Lahanas, N.~E.~Mavromatos and D.~V.~Nanopoulos,
  Phys.\ Lett.\  B {\bf 642} (2006) 179
  [arXiv:hep-th/0605181].
G.~A.~Diamandis, B.~C.~Georgalas, N.~E.~Mavromatos and E.~Papantonopoulos,
  Int.\ J.\ Mod.\ Phys.\  A {\bf 17} (2002) 4567
  [arXiv:hep-th/0203241];
  G.~A.~Diamandis, B.~C.~Georgalas, N.~E.~Mavromatos, E.~Papantonopoulos and I.~Pappa,
  Int.\ J.\ Mod.\ Phys.\  A {\bf 17} (2002) 2241
  [arXiv:hep-th/0107124];

\bibitem{lmn} A.~B.~Lahanas, N.~E.~Mavromatos and D.~V.~Nanopoulos,
  Phys.\ Lett.\  B {\bf 649}, 83 (2007)
  [arXiv:hep-ph/0612152].

\bibitem{aemn} G.~Amelino-Camelia, J.~R.~Ellis, N.~E.~Mavromatos and D.~V.~Nanopoulos,
  Int.\ J.\ Mod.\ Phys.\  A {\bf 12} (1997) 607
  [arXiv:hep-th/9605211].


\bibitem{gravanis} E.~Gravanis and N.~E.~Mavromatos,
Phys.\ Lett.\ B {\bf 547}, 117 (2002)
[arXiv:hep-th/0205298];
N.~E.~Mavromatos,
arXiv:hep-th/0210079 (published in {\it
Beyond the Desert, Oulu 2002 (Finland)} (ed. H.V. Klapdor-Kleingrothaus,
IoP 2003)), 3.

\bibitem{schmid} C.~Schmidhuber,
  Nucl.\ Phys.\ B {\bf 404}, 342 (1993)
  [arXiv:hep-th/9212075];
 C.~Schmidhuber and A.~A.~Tseytlin,
  Nucl.\ Phys.\ B {\bf 426}, 187 (1994)
  [arXiv:hep-th/9404180].



\bibitem{szabo} N.~E.~Mavromatos and R.~J.~Szabo,
  Phys.\ Rev.\  D {\bf 59} (1999) 104018
  [arXiv:hep-th/9808124].

\bibitem{emnmdr} J.~R.~Ellis, N.~E.~Mavromatos and D.~V.~Nanopoulos,
  Gen.\ Rel.\ Grav.\  {\bf 32} (2000) 127
  [arXiv:gr-qc/9904068].

\bibitem{polchinski} J.~Polchinski,
{\it String theory}, Vol. 2 (Cambridge University Press, 1998);
J.~H.~Schwarz,
arXiv:hep-th/9907061.

\bibitem{strominger} A.~Strominger,
  Phys.\ Lett.\  B {\bf 383} (1996) 44
  [arXiv:hep-th/9512059].

\bibitem{emw} J.~Ellis, N.~E.~Mavromatos and M.~Westmuckett,
Phys.\ Rev.\ D {\bf 70}, 044036 (2004)
[arXiv:gr-qc/0405066];
 Phys.\ Rev.\ D {\bf 71}, 106006 (2005)
[arXiv:gr-qc/0501060].




\bibitem{kmw} I.~I.~Kogan, N.~E.~Mavromatos and J.~F.~Wheater,
  Phys.\ Lett.\  B {\bf 387} (1996) 483
  [arXiv:hep-th/9606102];
I.~I.~Kogan and N.~E.~Mavromatos,
  Phys.\ Lett.\  B {\bf 375} (1996) 111
  [arXiv:hep-th/9512210].

\bibitem{horizon} J.~R.~Ellis, N.~E.~Mavromatos and D.~V.~Nanopoulos,
  Phys.\ Rev.\  D {\bf 62} (2000) 084019
  [arXiv:gr-qc/0006004].


\bibitem{rainbow} J.~Magueijo and L.~Smolin,
  Class.\ Quant.\ Grav.\  {\bf 21} (2004) 1725
  [arXiv:gr-qc/0305055].

\bibitem{volkov} J.~R.~Ellis, N.~E.~Mavromatos, D.~V.~Nanopoulos and G.~Volkov,
  Gen.\ Rel.\ Grav.\  {\bf 32} (2000) 1777
  [arXiv:gr-qc/9911055].

\bibitem{sarben} N.~E.~Mavromatos and S.~Sarkar,
  Phys.\ Rev.\  D {\bf 74} (2006) 036007
  [arXiv:hep-ph/0606048];
  Phys.\ Rev.\  D {\bf 72} (2005) 065016
  [arXiv:hep-th/0506242].

\bibitem{alfaro}
  R.~Gambini and J.~Pullin,
  Phys.\ Rev.\  D {\bf 59} (1999) 124021
  [arXiv:gr-qc/9809038];
J.~Alfaro, H.~A.~Morales-Tecotl and L.~F.~Urrutia,
  Phys.\ Rev.\  D {\bf 65} (2002) 103509
  [arXiv:hep-th/0108061];
  L.~Gonzalez-Mestres,
  arXiv:physics/9712056;
  arXiv:physics/9704017;
  arXiv:astro-ph/9505117.

\bibitem{ems} J.~R.~Ellis, N.~E.~Mavromatos and A.~S.~Sakharov,
  Astropart.\ Phys.\  {\bf 20} (2004) 669
  [arXiv:astro-ph/0308403];
J.~R.~Ellis, N.~E.~Mavromatos, D.~V.~Nanopoulos and A.~S.~Sakharov,
  Nature {\bf 428} (2004) 386
  [arXiv:astro-ph/0309144];
  Int.\ J.\ Mod.\ Phys.\  A {\bf 19} (2004) 4413
  [arXiv:gr-qc/0312044].





 \bibitem{crab} T.~Jacobson, S.~Liberati and D.~Mattingly,
  Nature {\bf 424} (2003) 1019
  [arXiv:astro-ph/0212190].


\bibitem{grb} G.~Amelino-Camelia, J.~R.~Ellis, N.~E.~Mavromatos, D.~V.~Nanopoulos and S.~Sarkar,
  Nature {\bf 393} (1998) 763
  [arXiv:astro-ph/9712103];
J.~R.~Ellis, K.~Farakos, N.~E.~Mavromatos, V.~A.~Mitsou and D.~V.~Nanopoulos,
  Astrophys.\ J.\  {\bf 535} (2000) 139
  [arXiv:astro-ph/9907340];
J.~R.~Ellis, N.~E.~Mavromatos, D.~V.~Nanopoulos and A.~S.~Sakharov,
  Astron.\ Astrophys.\  {\bf 402} (2003) 409
  [arXiv:astro-ph/0210124].

\bibitem{sakellariadou}
  N.~Mavromatos and M.~Sakellariadou,
  arXiv:hep-th/0703156, Phys. Lett. B in press.

\bibitem{bekenstein}  J.~D.~Bekenstein,
  Phys.\ Rev.\  D {\bf 70}, 083509 (2004)
  [Erratum-ibid.\  D {\bf 71}, 069901 (2005)].

\bibitem{skordis} C.~Skordis, D.~F.~Mota, P.~G.~Ferreira and C.~Boehm,
  Phys.\ Rev.\ Lett.\  {\bf 96} (2006) 011301
  [arXiv:astro-ph/0505519].


\bibitem{bmp} J.~Bernabeu, N.~E.~Mavromatos and J.~Papavassiliou,
  Phys.\ Rev.\ Lett.\  {\bf 92}, 131601 (2004)
  [arXiv:hep-ph/0310180].


\bibitem{noncom}
  S.~M.~Carroll, J.~A.~Harvey, V.~A.~Kostelecky, C.~D.~Lane and T.~Okamoto,
  Phys.\ Rev.\ Lett.\  {\bf 87} (2001) 141601
  [arXiv:hep-th/0105082].

\bibitem{toumbas} N.~Seiberg, L.~Susskind and N.~Toumbas,
  JHEP {\bf 0006} (2000) 044
  [arXiv:hep-th/0005015].

\bibitem{yoneya} T.~Yoneya,
  Prog.\ Theor.\ Phys.\  {\bf 103} (2000) 1081
  [arXiv:hep-th/0004074].



  \bibitem{genus}
  T.~Kuroki and S.~J.~Rey,
  Phys.\ Lett.\  B {\bf 499} (2001) 158
  [arXiv:hep-th/0007055].




\bibitem{vacaru} S.~I.~Vacaru,
  arXiv:hep-th/0211068;
  arXiv:hep-th/0310132.





\bibitem{grillo}
R.~Aloisio, A.~Galante, A.~F.~Grillo, S.~Liberati, E.~Luzio and F.~Mendez,
  Phys.\ Rev.\  D {\bf 74} (2006) 085017
  [arXiv:gr-qc/0607024];
  Phys.\ Rev.\  D {\bf 73} (2006) 045020
  [arXiv:gr-qc/0511031].


  \bibitem{bogosl}
  G.~Y.~Bogoslovsky,
  arXiv:0706.2621 [gr-qc].
\bibitem{smolstrings} J.~Magueijo and L.~Smolin,
  Phys.\ Rev.\  D {\bf 71} (2005) 026010
  [arXiv:hep-th/0401087].


\end{thebibliography}
\end{document}